\documentclass[12pt,x11names]{article}
\pdfoutput=1
\usepackage{amssymb,amsmath}
\usepackage{epsfig}
\usepackage{epstopdf}
\usepackage{latexsym}
\usepackage{graphicx}
\usepackage{booktabs}
\usepackage{bbm}
\usepackage{enumitem}
\usepackage[numbers,compress]{natbib}
\usepackage{float}
\numberwithin{equation}{section}
\usepackage{lmodern}

\usepackage[T1]{fontenc}

\usepackage[margin=20pt,small]{caption}
\usepackage{subcaption}

\usepackage[toc]{appendix}

\usepackage{overpic}
\usepackage{tikz}
\usetikzlibrary{decorations.pathreplacing}

\usepackage{array}
\usepackage[all]{xy}

\DeclareCaptionType{equ}[Diagram][]

\usepackage{xcolor}
\usepackage{datetime}

\newcommand*{\boxedcolor}{red}
\makeatletter
\renewcommand{\boxed}[1]{\textcolor{\boxedcolor}{%
  \fbox{\normalcolor\m@th$\displaystyle#1$}}}
\makeatother

\definecolor{cardinal}{rgb}{0.6,0,0}
\definecolor{darkgreen}{rgb}{0,0.5,0}
\definecolor{golden}{rgb}{0.92, 0.7, 0}
\definecolor{midnight}{rgb}{0, 0, 0.5}
\definecolor{darkblue}{rgb}{0.2, 0, 0.8}

\def\Re{{\rm Re}} \def\Im{{\rm Im}}

\newcommand{\arccosh}{\text{arccosh}\,}

\usepackage{staves}

\newcommand{\bb}[1]{\mathbb{#1}}
\newcommand{\dd}{\mathrm{d}}

\newcommand{\e}{\mathrm{e}}
\newcommand{\rmi}{i}
\newcommand{\rme}{\mathrm{e}}
\newcommand{\rmd}{\mathrm{d}}
\newcommand{\w}{\wedge}

\newcommand{\f}[2]{\frac{#1}{#2}}
\newcommand{\R}{\mathbf{R}}

\newcommand{\vol}{\text{vol}}
\renewcommand{\Re}{\text{Re}}
\renewcommand{\Im}{\text{Im}}

\newcommand{\Tr}{\text{Tr}~}

\newcommand{\SU}{\mathop{\rm SU}}
\newcommand{\SO}{\mathop{\rm SO}}

\newcommand{\SL}{\mathop{\rm SL}}
\newcommand{\U}{\mathop{\rm {}U}}

\def\Tr{{\rm Tr}\,}

\def\SL{{\rm SL}}

\def\SO{{\rm SO}}

\def\SU{{\rm SU}}

\usepackage{enumerate}
\numberwithin{equation}{section}

\usepackage{color}
\definecolor{dark-gray}{gray}{0.20}
\definecolor{gray}{gray}{0.30}
\definecolor{light-gray}{gray}{0.80}
\definecolor{dark-red}{rgb}{0.7,0,0}
\definecolor{dark-green}{rgb}{0.1,0.4,0}
\definecolor{dark-blue}{rgb}{0.3,0.3,0.7}
\definecolor{light-blue}{rgb}{0.8,0.8,1}

\newcommand{\be}{\begin{equation}}
\newcommand{\ee}{\end{equation}}
\newcommand{\bea}{\begin{eqnarray}}
\newcommand{\eea}{\end{eqnarray}}
\newcommand{\Z}{\mathbf{Z}}
\renewcommand{\Re}{\text{Re}}
\renewcommand{\Im}{\text{Im}}

\newcommand{\ISO}{\text{ISO}}

\newcommand{\ssl}{\mathfrak{sl}}
\newcommand{\Ess}{\text{E}_{7(7)}}

\newcommand{\ess}{\mathfrak{e}_{7(7)}}

\usepackage{hyperref}
\hypersetup{
     pdftoolbar=true,        
     pdfmenubar=true,        
     pdffitwindow=false,     
     pdfstartview={FitH},    
     pdfnewwindow=true,      
     colorlinks=true,       
     linkcolor=dark-blue,          
     citecolor=Green4,        
     filecolor=magenta,      
     urlcolor=DeepSkyBlue3,           
     linktocpage=true
}

\begin{document}  

\begin{titlepage}

\medskip
\begin{center} 
{\large \bf  Holographic 3d $\mathcal{N}=1$ Conformal Manifolds}

\bigskip
\bigskip
\bigskip
\bigskip

{\bf Nikolay Bobev,${}^{\text{\runictext{a}}}$ Fri\dh rik Freyr Gautason,${}^{\text{\runictext{b}}}$  and Jesse van Muiden${}^{\text{\runictext{c}}}$  \\ }
\bigskip
\bigskip
\bigskip
\bigskip
${}^{\text{\runictext{a}}}$Instituut voor Theoretische Fysica, KU Leuven \\
Celestijnenlaan 200D, B-3001 Leuven, Belgium
 \vskip 8mm
 ${}^{\text{\runictext{b}}}$University of Iceland, Science Institute,\\
 Dunhaga 3, 107 Reykjav\'ik, Iceland\\
  \vskip 8mm
${}^{\text{\runictext{c}}}$SISSA, Via Bonomea 265, 34136 Trieste, Italy
 
\bigskip 
\texttt{nikolay.bobev@kuleuven.be,~ ffg@hi.is,~jvanmuid@sissa.it} \\
\end{center}

\bigskip
\bigskip
\bigskip
\bigskip
\bigskip

\begin{abstract} 

\noindent We construct and study several examples of continuous families of AdS$_4$ $\mathcal{N}=1$ solutions of four-dimensional maximal gauged supergravity. These backgrounds provide the holographic descriptions of conformal manifolds of the dual  3d $\mathcal{N}=1$ SCFTs. The solutions we study can be uplifted to type IIB supergravity where they arise from D3-branes wrapping an $S^1$ with an S-duality twist. We find the spectrum of low lying operators in the 3d $\mathcal{N}=1$ SCFTs as a function of the exactly marginal coupling and discuss the structure of the corresponding superconformal multiplets. Using string theory techniques we also study additional examples of continuous families of $\mathcal N = 1$ AdS$_4$ vacua in type IIB and massive type IIA supergravity.

\end{abstract}

\noindent

\end{titlepage}


\setcounter{tocdepth}{2}
\tableofcontents

\section{Introduction}
\label{sec:intro}

The existence of conformal manifolds for CFTs in $d>2$ dimensions is usually associated with supersymmetry and ``the power of holomorphy''. Indeed, to the best of our knowledge all  examples of $d>2$ conformal manifolds are for 3d $\mathcal{N}=2$ SCFTs and 4d SCFTs with at least $\mathcal{N}=1$ supersymmetry.\footnote{See \cite{Green:2010da,Komargodski:2020ved,Perlmutter:2020buo} for a recent summary of known examples and general properties of ``holomorphic conformal manifolds'' and \cite{Baggio:2017mas} for a simple example of a non-trivial 3d $\mathcal{N}=2$ conformal manifold.} Moreover, it was shown in \cite{Cordova:2016xhm} that there cannot be any supersymmetric exactly marginal couplings for unitary SCFTs in $d>4$ or for 3d SCFTs with $\mathcal{N}\geq 3$ supersymmetry. This state of affairs naturally leads to the question whether there are any $d>2$ non-supersymmetric or 3d $\mathcal{N}=1$ conformal manifolds. Some general aspects of the properties of non-supersymmetric conformal manifolds for CFTs in $d>2$ were discussed recently in \cite{Bashmakov:2017rko,Behan:2017mwi,Hollands:2017chb,Sen:2017gfr}. In these papers various obstructions for the existence of exactly marginal operators and non-trivial constraints on the CFT data were studied. Given the lack of explicit examples, it is certainly of great interest to either find examples of CFTs for which these obstructions are avoided without ``the power of holomorphy'' and there are indeed exactly marginal operators, or to rule out their existence. A powerful tool to study the dynamics of strongly coupled CFTs is provided by the AdS/CFT correspondence. It is therefore natural to look for explicit examples of ``non-holomorphic'' conformal manifolds using top-down constructions rooted in string theory. The goal of this paper is to discuss a number of such examples for 3d $\mathcal{N}=1$ SCFTs by using their dual AdS$_4$ supergravity description.

The workhorse we will use in our holographic explorations of 3d $\mathcal{N}=1$ conformal manifolds is 4d $\mathcal{N}=8$ gauged supergravity. As summarized in \cite{deWit:2007kvg} this supergravity theory admits a number of distinct gaugings of subgroups of the $\Ess$ duality group of the ungauged supergravity theory constructed in \cite{Cremmer:1979up}. We will be mainly interested in two different gaugings that involve the non-compact groups  $[\SO(6)\times \SO(1,1)]\ltimes \mathbf{R}^{12}$ and $[\SO(6)\times \SO(2)]\ltimes \mathbf{R}^{12}$. As discussed in \cite{Inverso:2016eet}, see also \cite{Berman:2021ynm}, these two maximal supergravity theories can be uplifted to type IIB supergravity on $S^1\times \tilde{S}^5$ where $\tilde{S}^5$ denotes a squashed five-sphere. The construction is non-geometric since it involves a twist by an element of $\SL(2,\mathbf{Z})$ along the $S^1$. When the gauge group involves the $\SO(2)$ factor this so-called S-fold twist is by a compact generator of $\SL(2,\mathbf{Z})$, while for the $\SO(1,1)$ gauging one has to use a non-compact generator.

The $[\SO(6)\times \SO(1,1)]\ltimes \mathbf{R}^{12}$ gauged supergravity and its AdS$_4$ vacua have been the subject of recent interest and a number of non-trivial solutions were found and analyzed, see \cite{Guarino:2019oct,Guarino:2020gfe,Giambrone:2021zvp,Guarino:2021kyp,Bobev:2021yya,Guarino:2021hrc,Cesaro:2021tna}. Notably, several families of supersymmetric vacua were constructed in this supergravity theory. In particular, a two-parameter family of $\mathcal{N}=2$ AdS$_4$ solutions dual to a 3d $\mathcal{N}=2$ conformal manifold were found in \cite{Bobev:2021yya}, see also \cite{Guarino:2020gfe,Arav:2021gra}. It was also pointed out in \cite{Guarino:2020gfe,Guarino:2021kyp,Guarino:2021hrc} that there is a two-parameter family of $\mathcal{N}=1$ $\U(1)\times \U(1)$-invariant AdS$_4$ solutions of this supergravity theory. We briefly review these results below and emphasize that this family of solutions is holographically dual to a rare example of a 3d $\mathcal{N}=1$ conformal manifold. Building on the results of \cite{Guarino:2020gfe} we also compute the masses of all 4d supergravity modes for this family of AdS$_4$ vacua and organize them into superconformal multiplets of the dual 3d $\mathcal{N}=1$ SCFT. Motivated by the discussion in \cite{Bobev:2019jbi}, we point out that this construction can be generalized by replacing the internal $S^5$ with a toric Sasaki-Einstein manifold. This in turn leads to a large space of 3d $\mathcal{N}=1$ conformal manifolds that are associated with the 4d $\mathcal{N}=1$ quiver gauge theories obtained by placing D3-branes on the cone over toric Sasaki-Einstein spaces.

Encouraged by this observation we also study the $[\SO(6)\times \SO(2)]\ltimes \mathbf{R}^{12}$ gauged supergravity theory and its AdS$_4$ vacua. We find two distinct one-parameter families of $\mathcal{N}=1$ supersymmetric AdS$_4$ solutions. The first family has $\SO(3)$ symmetry at the origin of the conformal manifold which is broken to $\U(1)$ as one turns on the exactly marginal coupling.\footnote{This solution as well as its continuous deformations were independently found and discussed in the very recent work~\cite{Berman:2021ynm}.} The other family has only $\U(1)$ symmetry along the entire conformal manifold. In both cases we find that at least one point on the conformal manifolds can be found as a solution of a $\mathbf{Z}_2\times\mathbf{Z}_2\times\mathbf{Z}_2$ truncation of the $[\SO(6)\times \SO(2)]\ltimes \mathbf{R}^{12}$ gauged supergravity theory with 7 chiral superfields. For both of these families of AdS$_4$ solutions we explicitly compute the masses of all 4d supergravity fields and organize them in superconformal multiplets of the dual 3d $\mathcal{N}=1$ SCFT. Most of these multiplets are long and unprotected and the conformal dimensions exhibit interesting dependence on the exactly marginal coupling. Similarly to the discussion in \cite{Giambrone:2021zvp} we argue that the conformal manifolds in both of these cases are compact, i.e. they have the topology of $S^1$. In addition, we find 13 other AdS$_4$ non-supersymmetric vacua of this gauged supergravity theory which are perturbatively unstable.

In all three examples of 3d $\mathcal{N}=1$ conformal manifolds described above we observe a very similar structure in part of the superconformal multiplet spectrum.  The supergravity solutions suggest that at the origin of the conformal manifold there is a certain flavor symmetry G$_F$ which is broken to its Cartan subalgebra by the exactly marginal deformation. The exactly marginal scalar operators sit in a 3d $\mathcal{N}=1$ superconformal multiplet $L_2[2,0]$ with $\Delta=2$ scalar conformal primary. The $L_2[2,0]$ multiplet is in the adjoint representation of G$_F$ and at the points on the conformal manifold with enhanced flavor symmetry there is non-trivial multiplet rearrangements. This is similar in spirit to the superconformal multiplet re-organization along holomorphic conformal manifolds discussed in \cite{Green:2010da}. Moreover, we find that in all three examples there is a particular $L_1[3,1]$ multiplet with a spin-1 conformal primary which is not protected but nevertheless has operators with half-integer dimensions that do not depend on the marginal coupling. While we do not have a field theory understanding of this peculiar pattern of superconformal multiplets we observe that the $\SU(3)$ invariant $\mathcal{N}=1$ solution of the ISO(7) maximal gauged supergravity found in \cite{Guarino:2015qaa} has exactly these multiplets in its 4d supergravity spectrum. This leads to the natural conjecture that this supergravity solution also admits a two-dimensional space of continuous deformations dual to a 3d $\mathcal{N}=1$ conformal manifold. We test this proposal explicitly and find that it is not realized due to the structure of the cubic terms in the supergravity potential which do not allow for a flat direction.

There is another method to find explicit examples of conformal manifolds in top-down holography. One can start by a given AdS$_4$ vacuum of string or M-theory which is dual to a CFT with a continuous flavor symmetry that has a two-dimensional Cartan subalgebra (or a three-dimensional one in M-theory). If this symmetry is geometrically realized as an isometry of the internal manifold then one can apply the TsT transformation of Lunin-Maldacena \cite{Lunin:2005jy} to construct a family of AdS$_4$ solutions that are dual to a subspace of the conformal manifold in the dual CFT, see \cite{Imeroni:2008cr} for a review and \cite{Bobev:2021gza} for some recent examples of 3d $\mathcal{N}=2$ conformal manifolds constructed by this method and a more exhaustive list of references. If the TsT transformation is applied to an AdS$_4$ vacuum with $\mathcal{N}=1$ supersymmetry the result will be a non-holomorphic conformal manifold.\footnote{One can also imagine a situation where the seed vacuum solution has more than $\mathcal{N}=1$ supersymmetry but the TsT transformation breaks that to $\mathcal{N}=1$.} In the discussion below we identify three $\mathcal{N}=1$ supersymmetric vacua of massive type IIA supergravity that could be deformed in this way and briefly comment on their properties.

We proceed in the next Section with a short summary of the properties of the family of AdS$_4$  $\mathcal{N}=1$ solutions found in \cite{Guarino:2020gfe} and discuss the operator spectrum and properties of the conformal manifold of the dual 3d $\mathcal{N}=1$ SCFTs. We continue in Section~\ref{sec:AdS4FFG} with the presentation of the main new results in this work. First we outline the construction of two families of $\mathcal{N}=1$ vacua of 4d maximal supergravity with an $[\SO(6)\times \SO(2)]\ltimes \mathbf{R}^{12}$ gauging. We then proceed to study the spectrum of 4d supergravity excitations around these families of solutions and discuss the properties of the dual 3d $\mathcal{N}=1$ conformal manifolds. In Section~\ref{sec:other3dN1} we argue that several other examples of supersymmetric AdS$_4$ supergravity solutions with $\mathcal{N}=1$ supersymmetry also allow for continuous deformations that should be dual to 3d $\mathcal{N}=1$ conformal manifolds. We conclude with a short discussion in Section~\ref{sec:discussion}. In the Appendix we present a list of the 13 new non-supersymmetric AdS$_4$ solutions of the $[\SO(6)\times \SO(2)]\ltimes \mathbf{R}^{12}$ gauged supergravity we found and comment on some of their properties.

\section{Conformal manifolds from AdS$_4$ S-fold vacua}
\label{sec:AdS4Sfolds}

A two-parameter family of $\mathcal{N}=1$ AdS$_4$ solutions of the 4d $\mathcal{N}=8$ $[\SO(6)\times \SO(1,1)]\ltimes \mathbf{R}^{12}$ gauged supergravity theory was found in \cite{Guarino:2020gfe}. These solutions can be uplifted to S-fold solutions in type IIB supergravity of the schematic form AdS$_4\times S^1 \times \tilde{S}^5$, see \cite{Guarino:2021kyp} for more details. The $\tilde{S}^5$ represents a squashed five-sphere and there is a non-trivial monodromy of the $\SL(2,\mathbf{Z})$ symmetry of string theory along the $S^1$. The solutions are specified in terms of the constant values of three ``axionic'' scalar fields, $\chi_i$, in the 4d supergravity theory which obey the constraint
\begin{equation}\label{eqn:sumchi}
\chi_1+\chi_2+\chi_3=0\,.
\end{equation}
The mass spectrum of the 70 scalar and the 28 spin-1 fields in the 4d supergravity theory was computed in \cite{Guarino:2020gfe}. Our goal here is to translate this mass spectrum into the spectrum of conformal dimensions of operators in the dual 3d $\mathcal{N}=1$ SCFT and organize them into superconformal multiplets. To this end we also need to calculate the mass spectrum for the spin-1/2 and spin-3/2 fermions in the supergravity theory. After doing this calculation explicitly we can employ the standard holographic dictionary summarized in Table~\ref{tbl-dims} to find the conformal dimensions of all operators and then use 3d $\mathcal{N}=1$  superconformal representation theory to organize all operators into superconformal multiplets, see Table~\ref{tbl-N1rep}. We summarize the results of this calculation below.
\begin{table}[t]
\renewcommand{\arraystretch}{1.4}
\begin{center}
\begin{tabular}{@{\extracolsep{15 pt}}l l c }
\toprule
  Spin & Dimension &   \\
\midrule
\noalign{\smallskip}
0 & $\Delta =\frac{3}{2}\pm \sqrt{\frac{9}{4}+m^2L^2}$ \\
${1\over 2}$ & $\Delta  =\frac{3}{2}+|mL|$ \\
$1$ & $\Delta =\frac{3}{2}\pm \sqrt{\frac{1}{4}+m^2L^2}$  \\
${3\over 2}$ & $\Delta =\frac{3}{2}+|mL|$ \\
\bottomrule
\end{tabular}
\caption{Dimensions of operators dual to fields of spin, $s$, and mass, $m$. }
\label{tbl-dims}
\end{center}
\end{table}
%

For general values of $\chi_i$ there is the energy-momentum tensor multiplet, and two $\U(1)$ current multiplets:
\begin{equation}
	A_1[\tfrac52,\tfrac32]\,, \qquad 2\times A_1[\tfrac32,\tfrac12]\,.
\end{equation}
These are protected multiplets in which the conformal dimensions are fixed. In addition, we find various long multiplets. There are $7$ long spin-$1$ multiplets 
\begin{equation}
	L_1[3,1]\,, \qquad 2\times L_1[1 +  \sqrt{\tfrac{16}{9} +  \tfrac{5\chi_i^2}{4}},1]\,, \quad i=1,2,3 \,.
\end{equation}
We also find $12$ long spin-$1/2$ multiplets
\begin{equation}\label{eq:Longs12SU3}
	\begin{aligned}
	&2\times	L_1[1 + \sqrt{\tfrac{109}{36} + \tfrac{5 \chi_i^2}{4}},\tfrac12]\,,\\
	&2\times	L_1[1 + \tfrac12 \sqrt{1+5(\chi_i - \chi_j)^2},\tfrac12]\,,
	\end{aligned}
\qquad
\begin{aligned}
	  i<j\,, \quad i,j=1,2,3\,.
\end{aligned}
\end{equation}
Finally, there are $16$ long spin-$0$ multiplets
\begin{equation}\label{eq:Longs0SU3}
\begin{aligned}
	&2\times L_2[2,0]\,,  \\
	&2\times L_2[1+\sqrt{\tfrac49 + 5 \chi_i^2 },0]\,,
\end{aligned}
\qquad 
\begin{aligned}
	&2\times L_2[1+\sqrt{6},0]\,,\\
	&2\times L_2[1+ \sqrt{\tfrac{4}{9} + \tfrac{5 \chi_i^2}{4} }]\,,
\end{aligned}
\qquad 
\begin{aligned}
	i=1,2,3\,.
\end{aligned}
\end{equation}
%
%
\begin{table}[t]
\renewcommand{\arraystretch}{1.2}
\begin{center}
\begin{tabular}{@{\extracolsep{2 pt}}l c c c }
\toprule
\noalign{\smallskip}
Name \hspace{-10 pt} &  Primary & Descendants & Unitarity bound \\
\noalign{\smallskip}
\midrule
\noalign{\smallskip}
Identity, $B_1[0,0]$ & $|0,0\rangle$ & - & $\Delta=0$    \\
Short scalar, $A_2[\tfrac{1}{2},0]$& $|\frac{1}{2},0\rangle$ & $|\frac{1}{2},\frac{1}{2}\rangle$ & $\Delta=\frac{1}{2}$    \\
Short spin, $A_1[s+1,s]$& $|s+1,s\rangle$ & $|s+\frac{3}{2},s+\frac{1}{2}\rangle$ & $\Delta=s+1$; $s>0$    \\
Long scalar, $L_2[\Delta,0]$ & $|\Delta,0\rangle$ & $|\Delta+\frac{1}{2},\frac{1}{2}\rangle$; $|\Delta+1,0\rangle$ & $\Delta>\frac{1}{2}$    \\
Long spin, $L_1[\Delta,s]$ & $|\Delta,s\rangle$ & $|\Delta+\frac{1}{2},s+\frac{1}{2}\rangle$; $|\Delta+\frac{1}{2},s-\frac{1}{2}\rangle$; $|\Delta+1,s\rangle$ & $\Delta>s+1$; $s>0$    \\
\bottomrule
\end{tabular}
\caption{The 3d $\mathcal{N}=1$ superconformal multiplets, see \cite{Cordova:2016emh} for more details . The first column indicates also the notation we use for each multiplet which differs from the one used in \cite{Cordova:2016emh}.} \label{tbl-N1rep}
\end{center}
\end{table}
%
\noindent
We note that in order to compile this spectrum we have removed 26 Goldstone scalars with $m^2L^2=0$ which are the result of the breaking of the 28-dimensional gauge group of the supergravity theory to $\U(1)^2$. In addition we had to remove 7 Goldstone spin-1/2 fermions with masses that are related to those of the spin-3/2 fermions by $m^2_{s=1/2}=4m^2_{s=3/2}$. These goldstini arise due to the breaking of the $\mathcal{N}=8$ supersymmetry to $\mathcal{N}=1$. The explicit masses for the 7 goldstini modes are
\begin{equation}
m^2_{s=1/2}L^2 = 16\,, \qquad  2\times m^2_{s=1/2}L^2 =\tfrac{64}{9}+5\chi_i^2\,, \qquad i=1,2,3\,.
\end{equation}
The two exactly marginal operators sit in the $L_2[2,0]$ multiplets in \eqref{eq:Longs0SU3}. Notice also the appearance of long multiplets which have operator dimensions that do not depend on the parameters $\chi_i$. The $L_1[3,1]$ multiplet is particularly interesting since the operators in it have half-integer conformal dimensions and also this multiplet appears in the spectrum of all examples of conformal manifolds in 4d maximal gauged supergravity that we discuss in this work.

At special values of $\chi_i$ there is symmetry enhancement and multiplet rearrangements. For $\chi_i=0$ the symmetry enhances from $\U(1)^2$ to $\SU(3)$. The multiplet rearrangement capturing this is given in the middle line of Equation (2.4) in \cite{Cordova:2016emh}. More specifically we find that the six multiplets on the second line of \eqref{eq:Longs12SU3} become six $L_1[\tfrac32,\tfrac12]$ multiplets which then reorganize as follows
\begin{equation}\label{eq:multrear}
	6\times L_1[\tfrac32,\tfrac12] \rightarrow 6\times A_1[\tfrac32,\tfrac12]  + 6\times L_2[2,0]\,.
\end{equation}
The six $A_1[\tfrac32,\tfrac12]$ short multiplets reflect the enhancement of the flavor symmetry to $\SU(3)$ which is accompanied by the appearance of six new ``naively marginal'' multiplets $L_2[2,0]$. This multiplet rearrangement is very similar to the flavor symmetry enhancement and rearrangements appearing in 3d $\mathcal{N}=2$ and 4d $\mathcal{N}=1$ conformal manifolds \cite{Green:2010da}. We note that in order to complete the states in the six $L_2[2,0]$ multiplets in \eqref{eq:multrear} we need to remember that at $\chi_i=0$ there are only 20 Goldstone bosons instead of the 26 present for general values of $\chi_i$. In other words, for $\chi_i=0$ six of the Goldstone scalars with $m^2L^2=0$  become physical modes dual to $\Delta=3$ operators that are the bottom components of the six $L_2[2,0]$ multiplets in \eqref{eq:multrear}. A very similar mechanism is at play when two of the parameters $\chi_i$ are equal, for instance $\chi_1=\chi_2$ (and thus $\chi_3=-2\chi_1$ due to \eqref{eqn:sumchi}). In this case the symmetry is enhanced to $\SU(2)\times \U(1)$ and two of the multiplets on the second line of \eqref{eq:Longs12SU3} become two $L_1[\tfrac32,\tfrac12]$ multiplets that undergo the same multiplet rearrangement as in \eqref{eq:multrear}.

We also note that if any of the marginal parameters takes the special values $\chi_i=\frac{1}{3}$ or $\chi_i=\frac{2}{3}$ the long scalar multiplets on the second line of \eqref{eq:Longs0SU3} become $L_2[2,0]$ multiplets. This could suggest the appearance of new branches of the conformal manifold at these special values of $\chi_i$. It will be most interesting to study in more detail the full potential of the 4d $\mathcal{N}=8$ $[\SO(6)\times \SO(1,1)]\ltimes \mathbf{R}^{12}$ gauged supergravity and determine whether indeed there are additional families of supersymmetric AdS$_4$ solutions emerging at these values of $\chi_i$. 

In Section~\ref{sec:other3dN1} we discuss a generalization of the families AdS$_4$ vacua described above. These solutions can be viewed as corresponding to exactly marginal deformations of the AdS$_4\times S^1 \times {M}^5$ vacua of IIB supergravity presented in \cite{Bobev:2019jbi} where ${M}^5$ is a toric Sasaki-Einstein manifold.

We want to stress that while the conformal manifold described above appears to be non-compact, the parameters $\chi_i$ have to be periodically identified which in fact leads to a compact conformal manifold in harmony with the conjecture in \cite{Perlmutter:2020buo}. We discuss this in more detail in Section~\ref{subsec:FS3period}. We also note in passing that the supergravity kinetic terms for the scalars $\chi_i$ should capture the Zamolodchikov metric on the conformal manifold. We have computed these kinetic terms in the supergravity Lagrangian and have checked that this metric is flat. We therefore conclude that the 3d $\mathcal{N}=1$ conformal manifold at hand has $T^2$ topology with a flat metric.

\section{New $\mathcal{N}=1$ AdS$_4$ vacua and conformal manifolds}
\label{sec:AdS4FFG}

In order to describe the construction of the new families of AdS$_4$ vacua of interest here we start with a brief review of 4d $\mathcal{N}=8$ supergravity. Maximally supersymmetric, 4d supergravity consists of the metric, eight gravitini, 28 vector fields, 56 spin-1/2 gaugini, and 70 scalar fields. The scalar fields parametrize the coset space $\Ess/(\SU(8)/\Z_2)$. When the theory is ungauged it exhibits a large duality symmetry given by $\Ess$ which acts non-linearly on the scalar fields and the scalar potential is trivial. Our focus here is on gauged supergravity \cite{deWit:2007kvg} where a certain subgroup of $\Ess$ is promoted to a local symmetry group. This gauging leads to a modification of the supergravity Lagrangian and most notably induces a potential for the scalar fields of the theory. The supergravity Lagrangian is encoded in the embedding tensor ${\Theta_{\bb M}}^\alpha$ where $\alpha$ transforms in the $\bf 133$ of $\Ess$ and ${\bb M}$ transforms in the $\bf 56$. We use the real $\ssl(8)$  basis of $\ess$ written as $56\times56$ matrices $t_\alpha$ (see \cite{Cremmer:1979up,Bobev:2020qev} for details). Using this we define the $X$-tensor\cite{deWit:2007kvg}
\be
{X_{\bb M \bb N}}^{\bb R} = {\Theta_{\bb M}}^\alpha {(t_\alpha)_{\bb N}}^{\bb R}\,,
\ee
which plays an important role in the construction of the scalar potential of the theory. The embedding tensor transforms in the ${\bf 912}$ representation of $\Ess$, but in this paper we will focus on gaugings where it is built from ${\bf 36}\oplus{\bf 36}'$ representations of $\ssl(8)$. In this way the $X$-tensor is written in terms of two symmetric $8\times8$  matrices $\theta$ and $\xi$ through \cite{Inverso:2016eet}
\be\label{embeddingtensor}
\begin{split}
{X_{[AB][CD]}}^{[EF]}&=-X_{[AB]}{}^{[EF]}{}_{[CD]}=-8\delta^{[E}_{[A}\theta_{B][C}\delta^{D]}_{F]}\,,\\
X^{[AB]}{}_{[CD]}{}^{[EF]}&=-X^{[AB][EF]}{}_{[CD]}=-8\delta^{[A}_{[C}\xi^{B][E}\delta^{F]}_{D]}\,,
\end{split}
\ee
where $A,B,\dots =1,\dots,8$. We have represented the ${\bf 56}$ index as a direct sum ${\bf 28}\oplus {\bf 28}'$ of two representations of $\ssl(8)$, each of which is written as an antisymmetric combination $[AB]$. For example, a single ${\bf 56}$-vector can be represented as $x_{\bb M} = (x_{[AB]},x^{[AB]})$.

We will discuss a number of distinct gaugings which can be described simultaneously using diagonal matrices of the form
\be\label{thetaandxitensors}
\theta=g\,\text{diag}(1,1,1,1,1,1,a,b)\,,\quad \xi =g\,\text{diag}(0,0,0,0,0,0,\tilde a,\tilde b)\,,
\ee
where $g\neq 0$ is the gauge coupling constant. The quadratic constraint that the embedding tensor must satisfy for a consistent supergravity theory implies that $a\tilde a= b\tilde b = 0$. In Table \ref{tableofgaugings} we list the different gauge groups of interest in this work and the corresponding value for the parameters $a$, $\tilde a$, $b$, and $\tilde b$.
\begin{table}[t]
\begin{center}
\begin{tabular}{@{\extracolsep{10 pt}}l c c c c}
\toprule
Gauge symmetry& $a$ & $b$&  $\tilde a$ & $\tilde b$ \\
\midrule
$\SO(8)$ & 1 & 1  & 0 & 0\\
$\ISO(7)_c$ & 1 & 0  & 0 & $1$\\
$[\SO(6)\times \SO(1,1)]\ltimes \R^{12}$ & 0 & 0  & $1$ & $-1$\\
$[\SO(6)\times \SO(2)]\ltimes \R^{12}$ & 0 & 0  & $1$ & $1$\\
\bottomrule
\end{tabular}
\caption{The different gauge groups studied in this paper and their corresponding parameters in \eqref{superpot}. }
\label{tableofgaugings}
\end{center}
\end{table}

Since our interest here is in the properties of AdS$_4$ solutions of these supergravity theories we focus on the metric-scalar part of the supergravity Lagrangian. Later we also discuss the spectrum of fluctuations around the AdS$_4$ solutions for all supergravity fields. To this end we use the formulae presented in \cite{Gallerati:2014xra} for the mass matrices of supergravity fields with different spin. 

The 70 scalar fields parametrize the coset $\Ess/(\SU(8)/\Z_2)$ and are specified in terms of a $56\times 56$ scalar vielbein $\cal V$. From this we can build the scalar metric 
\be
{\cal M} = {\cal V}\cdot {\cal V}^T\,,\qquad {\cal M}^{\bb M \bb R} {\cal M}_{\bb R\bb N} = \delta_{\bb N}^{\bb M}\,.
\ee
The Lagrangian can now be written as \cite{deWit:2007kvg}\footnote{The metric is rescaled when compared to \cite{deWit:2007kvg} as follows $g^\text{here}_{\mu\nu} = 2 g^\text{there}_{\mu\nu}$. }
\be
{\cal L} = \sqrt{-g}\Big( R + \f1{48}\Tr\big(\dd {\cal M}\cdot \dd {\cal M}^{-1}\big)  -V \Big)\,,
\ee
where the scalar potential is
\be
V= \f{1}{168}{\cal M}^{\bb M \bb P} {X_{\bb M \bb N}}^{\bb R}{X_{\bb P \bb Q}}^{\bb S}\Big( {\cal M}^{\bb N \bb Q}{\cal M}_{\bb R \bb S} + 7\delta_{\bb S}^{\bb N}\delta_{\bb R}^{\bb Q}\Big)
\ee
%

\subsection{The 14-scalar model}
\label{subsec:14scalar}

To construct explicit AdS$_4$ solutions we focus on a consistent truncation of the supergravity theory that contains only a subset of the scalar fields. In particular we focus on a 14-scalar model constructed by keeping only fields that are invariant with respect to a $\Z_2\times \Z_2\times \Z_2$ subgroup of $\Ess$. This model has been  discussed by a number of authors, but we follow the conventions in \cite{Guarino:2020gfe,Bobev:2020qev}. We keep our discussion general so that we can treat the different gaugings we are interested in. The scalar manifold of this model consists of seven commuting $\SL(2,\R)/\U(1)$ cosets. Each $\SL(2,\R)_i$ with $i=1,\dots,7$ is generated by a positive and negative root generator $\mathfrak{e}_i$ and $\mathfrak{f}_i$ satisfying $\Tr(\mathfrak{e}_i\mathfrak{f}_j) = 6\delta_{ij}$ together with a Cartan generator $\mathfrak{h}_i=[\mathfrak{e}_i,\mathfrak{f}_i]$. A simple way to parametrize the $\SL(2,\R)$'s is to use
\be\label{70bein14scal}
{\cal V}_i = \e^{\sqrt{2}\Re z_i\,\mathfrak{e}_i}\cdot \e^{\log\Im z_i\, \mathfrak{h}_i}\,.
\ee
With this parametrization the 14-scalar model can be cast in a standard ${\cal N}=1$ supergravity language. In particular the scalar kinetic terms are specified by the K\"ahler potential
\be
K = -\sum_i \log( 2\Im z_i)\,.
\ee
Similarly the scalar potential can be written in terms of a superpotential $W$ via
\be\label{eq:potdef}
V = \e^{K}\Big(K^{i\bar{\jmath}}D_iW D_{\bar{\jmath}}\overline{W}-3W\overline{W})\,,
\ee
where the K\"ahler covariant derivative is defined as $D f = \partial f + f \partial K$.

Using the embedding tensor \eqref{embeddingtensor} and the two matrices $\theta$ and $\xi$ in \eqref{thetaandxitensors} we find that the superpotential can be written as a sum of two terms 
\be\label{superpot}
W= W_0 + 2g(\tilde a z_4z_5z_6z_7+\tilde b + z_1z_2z_3(a+b z_4z_5z_6z_7))\,,
\ee
where
\be
W_0=2g\big(z_1z_5z_6+z_2z_4z_6+z_3z_4z_5+z_1z_4z_7+z_2z_5z_7+z_3z_6z_7\big)\,.
\ee
The supergravity truncation and the superpotential are invariant under the discrete symmetry group $S_4$ that acts by permuting the scalar fields $z_i$, see \cite{Bobev:2021yya} for more details. If $a=1$ and $\tilde a=0$, then this symmetry gets enhanced to PSL$(2,7)$.

\subsection{Flat directions}
\label{flatdirs}

An important focus of our discussion is the construction of a number of continuous families of AdS$_4$ solutions. In all 4d supergravity examples we consider these continuous families are generated from a ``seed'' solution as in \cite{Bobev:2021yya,Guarino:2021hrc}. The seed solutions we utilize are described by the scalar vielbein ${\cal V}_0$ constructed as in \eqref{70bein14scal} and is therefore a solution of the 14 scalar model.\footnote{The same approach for generating families of AdS$_4$ vacua can be applied also to seed solutions outside of the 14-scalar truncation. An exhaustive study of all AdS$_4$ 4d gauged supergravity vacua admitting such deformations is outside of the scope of this paper.} The continuous families of solutions are then given by a product
\be\label{Familyofsolutions}
{\cal V} =  {\cal V}_{\chi}\cdot{\cal V}_0\,,
\ee
where ${\cal V}_{\chi}$ is a ``rotation'' matrix that generates the flat direction $\chi$. If there are multiple such direction then the scalar vielbein ${\cal V}$ is constructed as a product of multiple rotation matrices, one for each flat direction. Recently \cite{Guarino:2021hrc} showed how to construct the rotation matrices ${\cal V}_{\chi}$ for the $\SO(6)\times \SO(2)$ and the $\SO(6)\times \SO(1,1)$ gauged supergravity when ${\cal V}_0$ exhibits continuous global symmetry. In short, for each Cartan generator of the continuous symmetry, there is a generator $\mathfrak{g}$ such that the element 
\be
{\cal V}_\chi = \e^{\chi\, \mathfrak{g}}\,,
\ee
generates a flat direction as in \eqref{Familyofsolutions}. If the symmetry group of ${\cal V}_0$ is non-Abelian then on a generic point on the manifold of solutions only the Cartan subgroup is preserved. We refer to \cite{Guarino:2021hrc} for a more detailed discussion on the explicit construction of the generator $\mathfrak{g}$. 

Several comments on this construction are in order. The family of AdS$_4$ vacua constructed by the procedure described above may or may not be solutions of the 14 scalar model used to find the seed solution. We will exhibit examples of both cases below. The procedure to construct families of solutions is in principle independent of supersymmetry. If the seed solution exhibits supersymmetry and the Killing spinors are not charged with respect to the symmetry broken by the vielbein ${\cal V}_0$ then the whole family of AdS$_4$ vacua will remain supersymmetric. We will discuss both supersymmetric and non-supersymmetric families of 4d supergravity AdS vacua below.

We note that the construction discussed above is not the only possible way a family of AdS$_4$ gauged supergravity vacua can arise. Indeed in \cite{Bobev:2021yya} a two-dimensional family of ${\cal N}=2$ AdS$_4$ solutions was constructed analytically. One of those directions was already uncovered in \cite{Guarino:2020gfe} and is of the simple ``rotation kind'' discussed here. However, a second continuous family was found in \cite{Bobev:2021yya} which does not have a simple form in terms of a rotation matrix. 

\subsubsection{S-fold conformal manifold revisited}
Let us discuss briefly how this works for the conformal manifold discussed in Section~\ref{sec:AdS4Sfolds}. In this case we look for a solution in the $[\SO(6)\times \SO(1,1)]\ltimes \R^{12}$ gauged supergravity. This implies $\tilde a=-\tilde b =1$ and $a=b=0$ in the superpotential \eqref{superpot}. The $\SU(3)$ invariant seed solution is given by \cite{Guarino:2020gfe} 
\be
{\bf z} = \Big( \f{i\sqrt{5}}{3},\f{i\sqrt{5}}{3},\f{i\sqrt{5}}{3},\f{1+i\sqrt{5}}{\sqrt{6}},\f{1+i\sqrt{5}}{\sqrt{6}},\f{1+i\sqrt{5}}{\sqrt{6}},\f{1+i\sqrt{5}}{\sqrt{6}} \Big)\,.
\ee
Since the Cartan subgroup of $\SU(3)$ is two-dimensional we can use the prescription above to construct a two-dimensional manifold of AdS$_4$ solutions connected to this one. Generic point on the manifold preserves $\U(1)^2$ continuous global symmetry. Remarkably, the flat directions are entirely contained in the 14 scalar model and can be generated by shifting the first three $z_a$ coordinates by $\chi_a$ where $a=1,2,3$ and imposing the condition \eqref{eqn:sumchi} \cite{Guarino:2020gfe}. The value of the scalar potential for this family of supersymmetric solutions is
\be
V = -\f{162 g^2}{25\sqrt{5}}\,.
\ee

\subsection{New ${\cal N}=1$ AdS$_4$ solutions}

As an illustration of the procedure above we proceed with the presentation of two one-parameter family of $\mathcal{N}=1$ supersymmetric AdS$_4$ solutions of the $[\SO(6)\times \SO(2)]\ltimes \R^{12}$ gauged supergravity theory. To this end we fix $a=b=0$ and $\tilde{a}=\tilde{b}=1$ in Table~\ref{tableofgaugings} and in the superpotential \eqref{superpot} and we look for the critical points of the supergravity potential of the 14-scalar model \eqref{eq:potdef}. This results in a system of coupled algebraic equations which in general do not admit analytic solutions. Through an extensive numerical search we have identified 15 critical points of the potential corresponding to AdS$_4$ solutions with different values of the cosmological constant. If a given critical point of the potential is also a critical point of the superpotential \eqref{superpot}, then the AdS$_4$ solution is supersymmetric. We find two such supersymmetric solution of the 14-scalar model that we discuss in some detail below. In addition there are 13 non-supersymmetric critical points which we present in Appendix~\ref{app:nonsusyAdS4}. It is notable that 9 of the 15 critical points have a continuous subgroup of the $\SO(6)$ symmetry of the 4d gauged supergravity preserved. For all of these 9 AdS$_4$ vacua we can use the procedure outlined in Section~\ref{flatdirs} to construct families of AdS$_4$ solutions with the same value of the cosmological constant. The other 6 critical points we find have no continuous symmetry and correspond to isolated AdS$_4$ solutions.

One of the AdS$_4$ solutions we find has $\mathcal{N}=1$ supersymmetry and $\SO(3)\subset\SU(3)\subset\SO(6)$ symmetry. The critical point of the potential is found at the following values for the 7 scalars $z_i$ described in Section~\ref{subsec:14scalar}
\be\label{752908}
{\bf z} =\Big(\frac{2+i \sqrt{5}}{3 \sqrt{3}},\frac{2+i \sqrt{5}}{3 \sqrt{3}},\frac{2+i \sqrt{5}}{3
   \sqrt{3}},\frac{-1+i \sqrt{5}}{108^{1/4}},\frac{-1+i \sqrt{5}}{108^{1/4}},\frac{-1+i \sqrt{5}}{108^{1/4}},\frac{5+i \sqrt{5}}{108^{1/4}}\Big)\,.
\ee
The value of the potential \eqref{eq:potdef} at this critical point is
\be
V = -\f{243 \sqrt{3}\,g^2}{25\sqrt{5}}\,.
\ee
This solution is a part of a one-parameter family of $\mathcal{N}=1$ solutions with the same value of the cosmological constant parametrized by a real parameter $\chi$. For $\chi \neq0$ the $\SO(3)$ symmetry is broken to $\U(1)$ specified by the following $\SO(6)$ generator 
\begin{equation}\label{eq:u1genso3}
\mathfrak{g}_{\rm U(1)}=\left(
\begin{array}{cccccc}
 0 & 0 & 0 & 1 & 0 & 0 \\
 0 & 0 & -1 & 0 & 0 & 0 \\
 0 & 1 & 0 & 0 & 0 & 0 \\
 -1 & 0 & 0 & 0 & 0 & 0 \\
 0 & 0 & 0 & 0 & 0 & 0 \\
 0 & 0 & 0 & 0 & 0 & 0 \\
\end{array}
\right)\,.
\end{equation}

The other $\mathcal{N}=1$ solution we find has $\U(1)\subset\SO(6)$ symmetry and is specified by the following values for the 7 scalar fields $z_i$
\be\label{758297}\begin{split}
z_1 &=\f{\kappa}{\sqrt{6}}\e^{\f{i \pi}{3}}\,, \qquad  z_2 = z_3 =\f{\sqrt{3+\kappa^2}}{4\sqrt{2}}+ i\f{\sqrt{-3+3\kappa^2}}{4\sqrt{2}}\,,\\
z_4 &= \frac{\kappa-2}{2 \sqrt{3} \sqrt[4]{2 \kappa^2-5}}+\frac{i \sqrt[4]{\frac{7 \kappa^2}{2}-8}}{\kappa}\,, \qquad z_5 =z_6 = \frac{ \sqrt{\kappa}}{\sqrt[4]{6}}e^{\frac{2 i \pi }{3}}\,, \\
z_7 &= \frac{\sqrt{\kappa^2+2} \sqrt[4]{2 \kappa^2-5}}{\sqrt{6} (2-\kappa)}+\frac{i \sqrt[4]{\frac{7 \kappa^2}{2}-8}}{\kappa}\,,
\end{split}\ee
where we have defined the real constant $\kappa$ as
\be
\kappa = \sqrt{\sqrt{13}-1}\,.
\ee
The value of the scalar potential is 
\be
V = -\f{16 g^2}{27}\sqrt{70+26\sqrt{13}}\,.
\ee
The $\U(1)$ subgroup of $\SO(6)$ that is preserved by this solution is specified by the following generator of $\SO(6)$
\begin{equation}\label{eq:u1genu1}
\hat{\mathfrak{g}}_{\rm U(1)}=\left(
\begin{array}{cccccc}
 0 & 0 & 0 & 0 & 0 & 0 \\
 0 & 0 & 0 & 0 & 0 & 0 \\
 0 & 0 & 0 & 0 & 1 & 0 \\
 0 & 0 & 0 & 0 & 0 & -1 \\
 0 & 0 & -1 & 0 & 0 & 0 \\
 0 & 0 & 0 & 1 & 0 & 0 \\
\end{array}
\right)\,.
\end{equation}
Once again, this solution is a part of a one-parameter family of $\mathcal{N}=1$ supersymmetric AdS$_4$ vacua with the same value of the potential, parametrized by $\chi$ and generated by the procedure described in Section~\ref{flatdirs}.

\subsection{The spectrum}
\label{subsec:spec}

For the two families of supersymmetric AdS$_4$ vacua described above we have computed the masses of all bosonic and fermionic fields of the 4d supergravity theory as a function of the marginal parameter $\chi$. We have then used the AdS/CFT dictionary in Table~\ref{tbl-dims} to compute the conformal dimensions of all dual CFT operates and then organized them in superconformal multiplets according to Table~\ref{tbl-N1rep}. These results are similar to the ones in Section~\ref{sec:AdS4Sfolds} and are summarized below.

\subsubsection{$\SO(3)$ $\mathcal{N}=1$ AdS$_4$ vacuum}
\label{subsubsec:so3}
  
For generic values of the marginal parameter $\chi$ we find the energy-momentum tensor and one flavor current multiplets:
\begin{equation}
	A_1[\tfrac52,\tfrac32]\,, \qquad A_1[\tfrac32,\tfrac12]\,.
\end{equation}
There are 5 long multiplets with a spin-$1$ primary
\begin{equation}
	2\times L_1[3,1]\,,\qquad L_1[1 + \tfrac{\sqrt{41}}{3} ,1]\,,\qquad   2\times L_1[1 +\tfrac{1}{3\sqrt{2}}\sqrt{77 + \tfrac{25}{3}\chi^2 \pm 5 \sqrt{1+\tfrac{350}{9}\chi^2}} ,1]\,.
\end{equation}
We find $13$ long multiplets with spin-$1/2$ primaries
\begin{equation}
\begin{aligned}
	L_1[1 + \tfrac12\sqrt{\tfrac{43}{3}},\tfrac12]\,,\qquad 2\times L_1[1+ \tfrac{\sqrt{41}}{2},\tfrac12]\,, \qquad 2 \times L[1 + \sqrt{\alpha_n},\tfrac12]\,, ~~n=1,2,3 \\
	2\times L_1[1 + \tfrac12\sqrt{\tfrac{43}{3} + \tfrac23 \left(\tfrac{10\chi}{3}\right)^2 },\tfrac12]\,,\qquad  2\times L_1[1 +  \tfrac12\sqrt{\tfrac{43}{3} +\tfrac23 \left(\tfrac{5\chi}{3}\right)^2},\tfrac12]\,,
\end{aligned} 
\end{equation}
where $\alpha_{n}$ are the roots of a cubic polynomial
\begin{equation}
4 \alpha _n^3-\big(\tfrac{50 \chi ^2}{9}+83\big) \alpha _n^2+\big(\tfrac{625 \chi ^4}{243}+\tfrac{21025 \chi ^2}{729}+\tfrac{1763}{4}\big) \alpha _n-\tfrac{15625 \chi ^6}{39366}+\tfrac{349375 \chi ^4}{78732}-\tfrac{106675 \chi ^2}{1944}-\tfrac{1681}{16}\,,
\end{equation}
and explicitly take the form
\begin{equation}
	\alpha_n = \frac{1}{4} +\frac{10}{9} w_n + \frac{10q_2}{243w_n} + \frac{q_3}{9}\,, \quad 	w_n =(-1)^{\frac{2(n-1)}{3}} \sqrt[3]{\tfrac{q_1}{3} + \tfrac19 \sqrt{ 9 q_1^2 - \tfrac19 \left(\tfrac{q_2}{3}\right)^3}}\,, 
\end{equation}
and
\begin{equation}
	\{q_1 ,q_2,q_3\} =\{50\chi^2 - 9^2  , \tfrac{175\chi^2}{2} + \tfrac{9^3}{3},  \tfrac{25\chi^2}{6}  + 60 \}\,.
\end{equation}
We also find $15$ long multiplets with a spin-$0$ primary:
\begin{equation}\label{Eq: long spin-0 multiplets}
\begin{aligned}
		2\times L_2[1 +  \tfrac{1}{3\sqrt{2}}\sqrt{93 + 12 \left(\tfrac{5\chi}{3}\right)^2 \pm 5 \sqrt{17^2 + 56\left(\tfrac{5\chi}{3}\right)^2}},0]\,, \qquad  L_2[\tfrac53,0] \,, \qquad  L_2[1 + \tfrac{\sqrt{89}}{3},0]\,,\\ 
		2\times L_2[1 +  \tfrac{1}{3\sqrt{2}}\sqrt{93 + 3\left(\tfrac{5\chi}{3}\right)^2 \pm 5 \sqrt{17^2 + 14 \left(\tfrac{5\chi}{3}\right)^2}},0]\,, \qquad  L_2[5,0] \,,\qquad L_2[1 + \sqrt{\gamma_n},0]\,,
\end{aligned}
\end{equation}
and a multiplet that contains the exactly marginal operator 
\begin{equation}
	L_2[2,0]\,.
\end{equation}
The values of $\gamma_{n=1,2,3}$ in \eqref{Eq: long spin-0 multiplets} are again roots of a cubic polynomial
\begin{equation}
81 \gamma _n^3-1998 \gamma _n^2+11153 \gamma _n-1236\,,
\end{equation}
and are 
\begin{equation}
\gamma_1 \approx0.113103\,,\qquad \gamma_2 \approx8.30122\,,\qquad \gamma_2 \approx 16.2523\,.
\end{equation}

When we organized the spectrum above we have removed the 27 massless Goldston scalars with arising from the breaking of the 4d supergravity gauge group to $\U(1)$. We have also removed the 7 spin-$1/2$ goldstini that arise from the breaking of $\mathcal{N}=8$ supersymmetry to $\mathcal{N}=1$. The masses of the goldstini are given by
\begin{equation}
	2\times m^2L^2= 16\,,\qquad m^2L^2= \tfrac{164}{9}\,, \qquad 2\times m^2L^2=  \tfrac{154}{9} + \tfrac{50}{27}\chi^2 \pm \tfrac{10}{9} \sqrt{1 + \tfrac{350}{9} \chi^2} \,.
\end{equation}

At the origin of the conformal manifold $\chi \rightarrow 0$ we find that the two long multiplets $L_1[1 + \sqrt{\alpha_2}, \tfrac12]$ become two $L_1[\tfrac32 , \tfrac12]$ multiplets, which then recombine as
\begin{equation}
2\times L_1[\tfrac32,\tfrac12] \rightarrow 2\times A_1[\tfrac32,\tfrac12]  + 2\times L_2[2,0]\,.
\end{equation}
As discussed around \eqref{eq:multrear} this multiplet recombination is associated with the flavor symmetry enhancement to $\SO(3)$ at the origin of the conformal manifold.
 
\subsubsection{$\U(1)$ $\mathcal{N}=1$ AdS$_4$ vacuum}
\label{subsubsec:u1}

\begin{figure}[h!]
\centering
\begin{subfigure}[b]{0.3\textwidth}
\centering
\includegraphics[width=\textwidth]{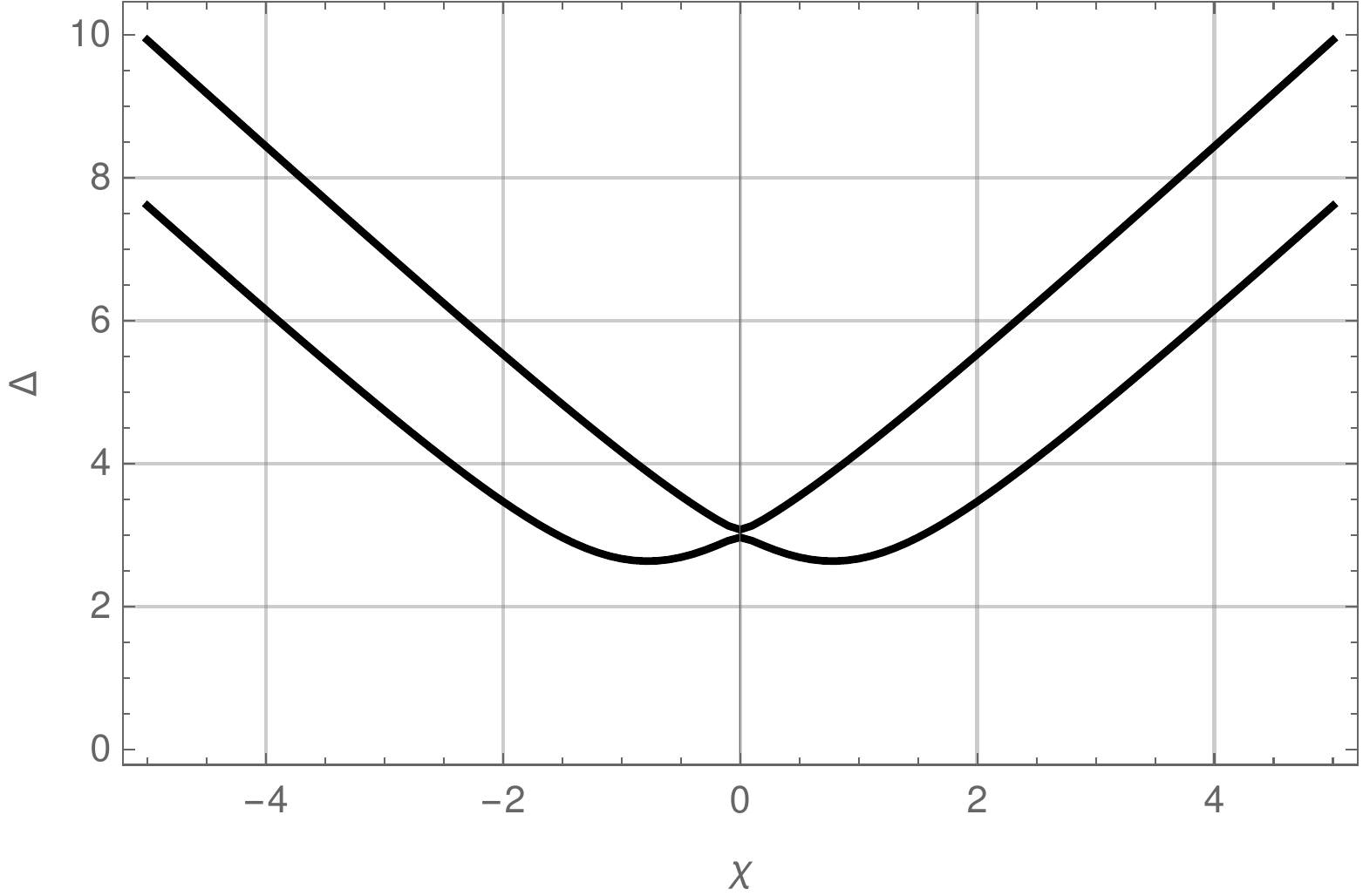}
\caption{$\zeta_a$}
\label{LongSpin1Plot}
\end{subfigure}
\begin{subfigure}[b]{0.3\textwidth}
\centering
\includegraphics[width=\textwidth]{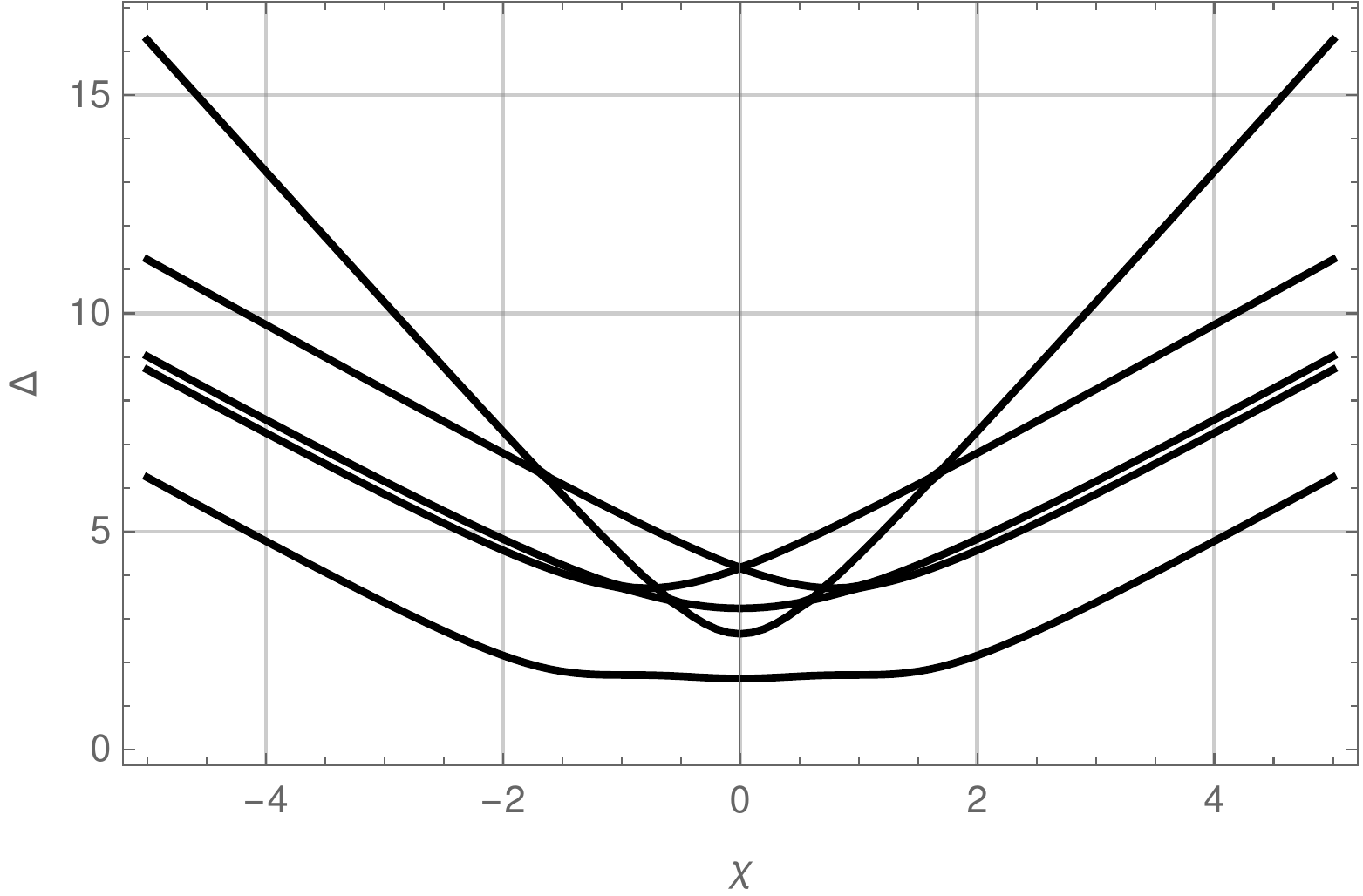}
\caption{$\eta_a$}
\label{LongSpin12Plot}
\end{subfigure}
\begin{subfigure}[b]{0.3\textwidth}
\centering
\includegraphics[width=\textwidth]{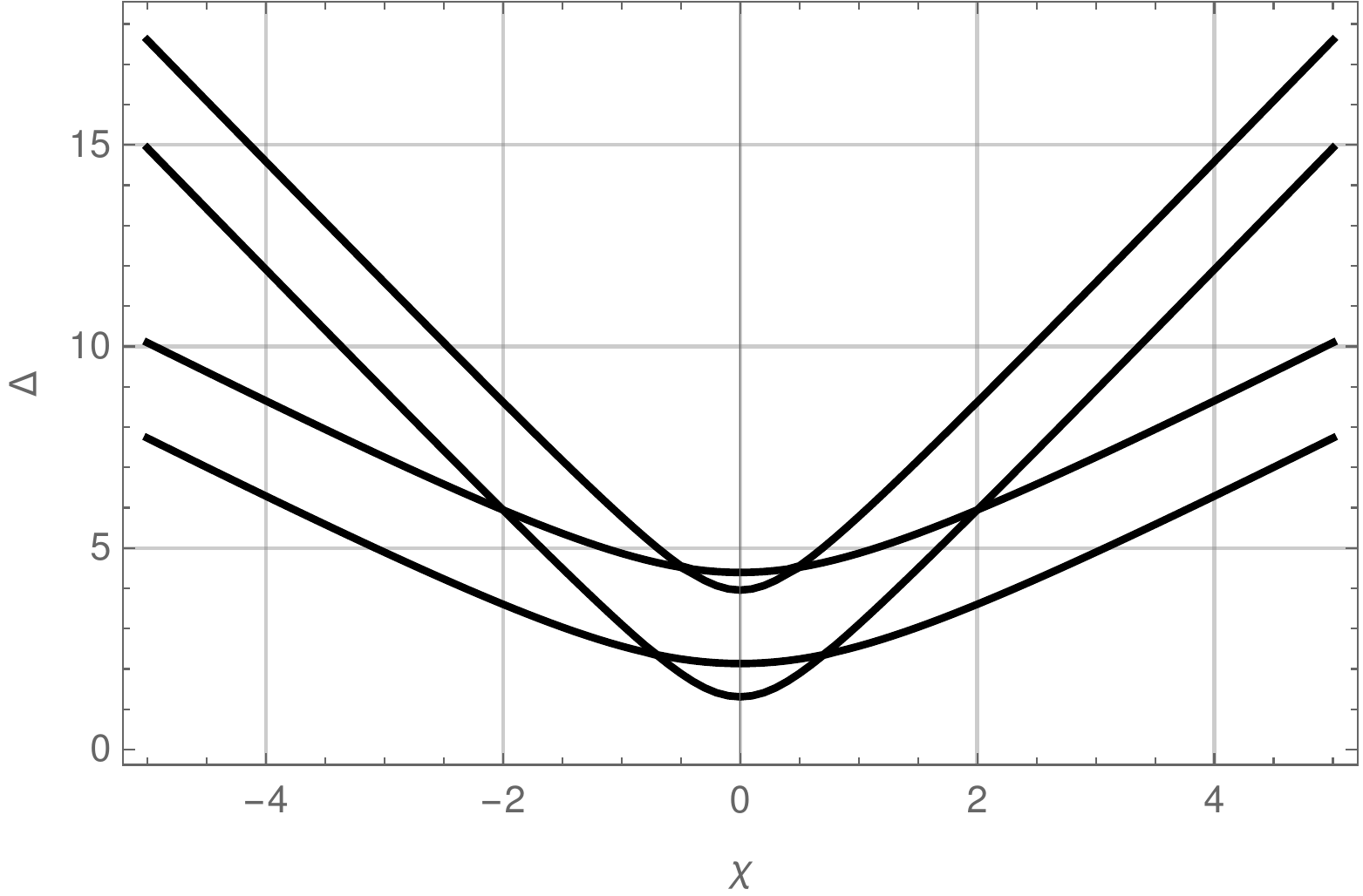}
\caption{$\omega_a$}
\label{LongSpin0Plot}
\end{subfigure}
\caption{Dimensions of primaries in the long multiplets which depend non-trivially on the marginal parameter $\chi$. From left to right we have dimensions of long spin-1, long spin-1/2, and long spin-0 multiplets.}
\end{figure}

We again have two short multiplets in the spectrum corresponding to the energy-momentum tensor and $\U(1)$ flavor current multiplets:
\begin{equation}
	A_1[\tfrac52,\tfrac32]\,, \quad A_1[\tfrac32,\tfrac12]\,.
\end{equation}
There are 7 long spin-1 multiplets in the spectrum. Three of them have conformal dimensions that do not depend on the parameter $\chi$. The dependence on $\chi$ of the other 4 multiplets can be found numerically and is presented in Figure \ref{LongSpin1Plot}
\begin{equation}
	L_1[3,1]\,, \quad L_1[\tfrac{3}{2}+\tfrac{\sqrt{13}}{2} ,1] \,, \quad L_1[1+\sqrt{\tfrac{11}{3}+\tfrac{\sqrt{13}}{3}},1] \,, \quad  2\times L_1[\zeta_a(\chi) ,1]\,, \qquad a=1,2\,.
\end{equation}
There are 13 spin-$1/2$ multiplets. Three of them have conformal dimensions that do not depend on the parameter $\chi$. The dependence on $\chi$ of the other 10 multiplets can be found numerically and is presented in Figure \ref{LongSpin12Plot}
\begin{equation}
\begin{split}
	& L_1[1+\sqrt{3+\tfrac{3\sqrt{13}}{4}},\tfrac{1}{2}]\,, \qquad L_1[1+\sqrt{\tfrac{29}{4}+\sqrt{13}},\tfrac{1}{2}] \,, \qquad L_1[1+\sqrt{\tfrac{20}{3}+\tfrac{13\sqrt{13}}{12}},\tfrac{1}{2}] \,,\\
	&   2\times  L_1[\eta_a(\chi) ,\tfrac{1}{2}]\,, \qquad a=1,\ldots, 5\,.
\end{split}	   
\end{equation}
We find also 15 spin-$0$ multiplets. One of them, $L_2[2,0]$, contains the exactly marginal operator. There are 6 more multiplets with dimensions that do not depend on $\chi$. The dependence on $\chi$ of the other 8 multiplets can be found numerically and is presented in Figure \ref{LongSpin0Plot}
\begin{equation}
\begin{split}
&L_2[2,0]\,, \qquad L_2[1+\sqrt{8+2\sqrt{13}},0]\,, \qquad L_2[1+\sqrt{1+r_n},0]\,, \quad n=1,\ldots,5 \\
& 2\times L_2[\omega_a(\chi),0]\,, \qquad a=1,\ldots, 4\,.
\end{split}
\end{equation}
The real numbers $r_n$ above do not depend on $\chi$ and have the approximate values
\begin{equation}\label{eq:rndef}
r_1 \approx -0.924201\,, \quad r_2\approx 0.978564\,,\quad  r_3 \approx 7.22535\,, \quad r_4 \approx 11.4676\,, \quad r_5 \approx 14.5036\,. 
\end{equation}
The $r_n$ can be computed numerically to high precision as roots of the following polynomial 
\begin{equation}
\begin{split}
&50544 + 35100 \,r_n - 202731\,r_n^2 + 286276\,r_n^3 - 6513\,r_n^4 - 329207 \,r_n^5 \\
 &\quad + 208011 \,r_n^6 - 52960 \,r_n^7 + 6577 \,r_n^8 - 393 \,r_n^9 + 9 \,r_n^{10}\,.
 \end{split}
\end{equation}
This polynomial has 8 real roots and the $r_n$ in \eqref{eq:rndef} are 5 of those roots.

As usual we should bear in mind that to organize the masses of the supergravity modes in the multiplets above we had to remove the $27$ Goldstone scalars with $m^2L^2=0$ as well as the $7$ spin-$1/2$ goldstini from the mass spectrum.

In all three examples of families of AdS$_4$ vacua dual to 3d $\mathcal{N}=1$ conformal manifolds discussed above the exactly marginal operators lie in the $L_2[2,0]$ superconformal multiplets which in turn transform in the adjoint representation of the flavor group on all points of the conformal manifold, including the ones with enhanced global symmetry. Moreover in all three examples we observe the presence of the curious $L_1[3,1]$ which is a singlet under the flavor group. As we discuss in Section~\ref{subsubsec:iso7} the presence of these multiplets in the spectrum of other supersymmetric AdS$_4$ vacua does not always guarantee the existence of exactly marginal deformations.

\subsection{Free energy and periodicity of $\chi$}
\label{subsec:FS3period}
After constructing the two families of supergravity solutions we can now employ the holographic dictionary to compute the $S^3$ free energy of the dual 3d $\mathcal{N}=1$ CFTs. To do this we also have to make use of the fact that all AdS$_4$ solutions we discussed above can be uplifted to type IIB supergravity where they describe the backreaction of $N$ D3-branes with an $\mathbf{R}^{1,3}\times S^1$ world volume and an $\SL(2,\mathbf{Z})$ S-duality twist on the $S^1$.

We start with the standard formula that relates the AdS length scale $L_\text{AdS}$ to the free energy \cite{Emparan:1999pm}
\be
{\cal F} = \f{\pi L_\text{AdS}^2}{2 G_N}\,,
\ee
where $G_N$ is the four-dimensional Newton constant. The latter is related to ten-dimensional quantities via
\be
\f{1}{16\pi G_N} = \f{2\pi}{(2\pi\ell_s)^8} \vol(S^5)\vol(S^1_\rho)\,,
\ee
where $\vol(S^5)\vol(S^1_\rho)$ is the volume of the internal space. Notice that for both $\SO(6)$ gaugings discussed in this paper the uplift to IIB supergravity involves an $S^5$ and a 1d space that is compactified by S-folding \cite{Inverso:2016eet,Berman:2021ynm}. For simplicity of notation we denote this sixth internal direction by $S^1_\rho$ even though it is not strictly speaking a geometric circle.
The four-dimensional Newton constant is most easily computed at the origin of the 4d supergravity scalar manifold even though that is not a critical point of the potential. The final answer for $G_N$ will of course not depend on the scalar configuration we use since the four-dimensional theory is written in Einstein frame. At the origin, the volume of the five-sphere is $\vol(S^5) = (2/g)^5 \pi^3$ where $g$ is the 4d gauge coupling constant and $2/g$ plays the role of the radius of $S^5$. In order to relate the string length $\ell_s$ to the number of D3-branes $N$, we use five-form flux quantization. Again we can perform the flux quantization at the origin of the 4d scalar manifold since the integer flux quanta cannot change continuously as a function of the supergravity scalars. The RR four-form is simply given by $C_4 = 4(2/g)^4 \omega_4$ where $\dd\omega_4$ is the volume form on unit-$S^5$. Flux quantization then results in
\be
N = \f{1}{(2\pi\ell_s)^4} \int \dd C_4 = \f{4}{\pi(\ell_s g)^4}\,, 
\ee
where we used that the volume of the unit-radius $S^5$ is $\pi^3$. 

Let us now write $\vol(S^1_\rho) = (2/g) \Delta \rho$ where $\rho$ is a coordinate parametrizing $S^1_\rho$ and $\Delta\rho$ is its range. The range for $\rho$ depends on the gauging in question. For the $\SO(1,1)$ case, it is given by $\Delta\rho = \arccosh(n/2)$ where the natural number $n$ specifies a hyperbolic element of $\SL(2,\mathbf{Z})$ which determines the particular S-duality twist at hand and, in accordance with its $\mathcal N=4$ and $\mathcal N=2$ cousins, has an interpretation as the Chern-Simons level of the gauge field in the dual 3d SCFT \cite{Inverso:2016eet,Assel:2018vtq,Bobev:2021yya}. For the $\SO(2)$ gauging the volume is $\Delta\rho = 2\pi( k + 1/m)$ where $k\in \bb{N}$ and $m=2,3,4,6$ \cite{Berman:2021ynm}. The value of $m$ specifies one of the 4 elliptic conjugacy classes of $\SL(2,\mathbf{Z})$ while the natural number $k$ determines a particular representative of $\SL(2,\mathbf{Z})$.

Combining these volume factors and using the result from flux quantization we obtain the four-dimensional Newtons constant
\be
\frac{1}{ G_N} = \f{g^2  N^2\Delta\rho}{2\pi}  \,.
\ee
The 3-sphere free energy is then 
\be
{\cal F} = \frac{3g^2  N^2\Delta\rho}{-2V^*}\,,
\ee
where we have replaced the AdS$_4$ length scale by the  value of the scalar potential at the critical point $V^* = -6/L_\text{AdS}^2$.

For the family of ${\cal N}=1$ AdS vacua in $\SO(1,1)$ gauged supergravity discussed in Section~\ref{sec:AdS4Sfolds} we recover the result obtained using five-dimensional supergravity \cite{Bobev:2019jbi,Bobev:2020fon}
\be
{\cal F} = \sqrt{\frac{5^5}{3^6}}\frac{N^2}{4}\arccosh \frac{n}{2}\,.
\ee
For the $\SO(3)$ invariant ${\cal N}=1$ AdS vacua in the $\SO(2)$ gauged supergravity we find \cite{Berman:2021ynm}
\be
{\cal F} = 
\sqrt{\frac{5^5}{3^9}} \frac{N^2}{2} 2\pi \left(k+\frac{1}{m}\right)
\ee
For the $\U(1)$ invariant ${\cal N}=1$ AdS vacua in $\SO(2)$ gauged supergravity we find the following free energy
\be
{\cal F} = \f{81 }{32\sqrt{70+26\sqrt{13}}}2\pi\Big( k + \f{1}{m}\Big)N^2\,.
\ee

Naively it seems that the conformal manifold dual to the family of supergravity solutions is non-compact. Indeed the range of $\chi$ is completely unrestricted in 4d supergravity. However, as explained in \cite{Giambrone:2021zvp,Guarino:2021kyp} the axionic scalars $\chi$ corresponds to a local coordinate transformation in ten dimensions. More concretely, a coordinate $\phi$ parametrizing an isometry direction on $S^5$ is shifted by the S-fold direction $S^1_{\rho}$ as follows
\be
\phi \to \hat\phi =\phi+\chi\rho\,.
\ee
This coordinate transformation is globally well defined only when $\chi = 2\pi p/\Delta\rho$ and $p\in\mathbb{Z}$ and therefore corresponds to the same ten-dimensional geometry. The parameter $\chi$ should therefore be thought of as being periodic with $\chi\sim \chi + 2\pi p/\Delta\rho$. This implies that the three conformal manifolds discussed in this paper are all compact.\footnote{The same holds for the non-supersymmetric families discussed in appendix \ref{app:nonsusyAdS4}.}

It is interesting to understand the implications of the compactness of $\chi$ for the spectrum computed in Section~\ref{subsec:spec}, in particular for the multiplets with conformal dimensions that depend non-trivially on $\chi$. If we restrict $\chi$ to its fundamental domain we must include new multiplets in the spectrum corresponding to shifting $\chi$ by its periodicity since the conformal dimensions are not $(2\pi /\Delta\rho)$-periodic. These infinite families of new multiplets should correspond to type IIB supergravity KK modes that arise from the KK spectrum on $S^1_\rho$. It would be very interesting to confirm this expectation by an explicit KK spectrum calculation as was done for another family of AdS$_4$ solutions in \cite{Giambrone:2021zvp}, see also \cite{Cesaro:2021tna}.

\section{Conformal manifolds for other AdS$_4$ solutions}
\label{sec:other3dN1}

The goal of this section is to consider several other examples of supersymmetric AdS$_4$ vacua arising from string and M-theory and to study whether they are dual to SCFTs with exactly marginal supersymmetric deformations.

\subsection{Looking for marginal deformations in 4d supergravity}

We start our discussion with the known supersymmetric vacua of the two other maximal gauged supergravity presented in Table~\ref{tableofgaugings}, namely the $\ISO(7)_c$ and $\SO(8)$ gauged supergravities.

\subsubsection{$\ISO(7)_c$}
\label{subsubsec:iso7}

As summarized in \cite{Bobev:2020qev} there are 7 known supersymmetric AdS$_4$ vacua of the maximal ISO(7) gauged supergravity theory. The spectrum of masses of the 4d supergravity fields around each of these vacua was presented in \cite{Bobev:2020qev} and organized into superconformal multiplets. For 6 of the vacua it is easy to check that there are no candidate exactly marginal operators in the 4d supergravity spectrum, i.e. there are no $L_2[2,0]$ superconformal multiplets in the spectrum. For the $\mathcal{N}=1$ $\SU(3)$ invariant AdS$_4$ vacuum however the story is more interesting. This solution was first constructed in \cite{Guarino:2015qaa} and can be found as a critical point of the potential and superpotential of the 14-scalar model described in Section~\ref{subsec:14scalar} for the following values of the 7 complex scalars
\be\label{eq:iso7su3vacz}
{\bf z} = \f14\Big( -1 + i\sqrt{15},-1 + i\sqrt{15},-1 + i\sqrt{15},\sqrt{3} + i\sqrt{5},\sqrt{3} + i\sqrt{5},\sqrt{3} + i\sqrt{5},\sqrt{3} + i\sqrt{5} \Big)\,.
\ee
The potential for this critical point takes the value
\be
V = -\f{48\sqrt{3}\, g^2}{25\sqrt{5}}\,.
\ee
As discussed in \cite{Bobev:2020qev} the 3d CFT operators dual to the 4d supergravity modes can be organized into the following multiplets. The stress-energy tensor and $\SU(3)$ flavor current multiplets
\begin{equation}
	A_1[\tfrac52,\tfrac32]\,, \qquad 8\times A_1[\tfrac32,\tfrac12]\,,
\end{equation}
as well as the following long multiplets
\begin{equation}\label{eq:iso7su3spec}
\begin{split}
& L_1[3,1]\,, \qquad 6\times L_1[\tfrac{7}{3},1]\,, \\
& 6\times L_1[1+\sqrt{\tfrac{109}{36}},\tfrac{1}{2}]\,, \\
& 8\times L_2[2,0]\,, \qquad 2\times L_2[1+\sqrt{6},0]\,, \qquad 12\times L_2[\tfrac{5}{3},0]\,. \\
\end{split}
\end{equation}
Notably, we observe the appearance of the $L_2[2,0]$ multiplet in the adjoint representation of the flavor group together with a single $L_1[3,1]$ multiplet. Based on the discussion of the spectrum for the families of supergravity solutions in Sections~\ref{sec:AdS4Sfolds} and \ref{subsec:spec}, this naively suggests that the SCFT dual to this AdS$_4$ supergravity vacuum admits exactly marginal deformations. The natural conjecture is that there are two such exactly marginal deformations which break the $\SU(3)$ flavor symmetry to its $\U(1)^2$ Cartan subalgebra. Unfortunately this expectation does not bear out. If such a family of supersymmetric AdS$_4$ vacua exists it should be a solution to the $\U(1)^2$ truncation of the $\ISO(7)_c$ described in \cite{Kim:2018cpz}, see also \cite{Guarino:2017pkw}. We have analyzed the properties of the potential in this model in detail and found that there are no continuous families of supersymmetric solutions. In particular we have studied the neighborhood of the location of the $\SU(3)$ critical point in \eqref{eq:iso7su3vacz} and have found no continuous deformation of this AdS$_4$ vacuum. To understand better the failure of the $L_2[2,0]$ multiplets in \eqref{eq:iso7su3spec} to lead to exactly marginal deformations we have also analyzed the cubic couplings in the superpotential in the $\U(1)^2$ invariant supergravity truncation. We find that the two $L_2[2,0]$ multiplets that are singlets under the $\U(1)^2$ symmetry have scalar operators with $\Delta=3$ but the three point functions of these operators do not vanish. This is manifested in the supergravity description by non-vanishing cubic terms for the corresponding massless supergravity scalars. This analysis provides an explanation for the absence of continuous families of AdS$_4$ solutions. From the supergravity perspective we have a massless scalar with a non-trivial cubic potential that lifts the flat direction. From the perspective of the dual CFT this non-vanishing cubic coupling results in a non-zero three-point function of the naively marginal operators which in turn means that this operator has a non-zero anomalous dimension at first order in  conformal perturbation theory and thus will not be exactly marginal \cite{Behan:2017mwi}.

\subsubsection{$\SO(8)$}

There are 5 supersymmetric AdS$_4$ vacua of the 4d maximal $\SO(8)$ gauged supergravity theory, see \cite{Comsa:2019rcz,Bobev:2019dik}. Out of these critical points only the $\U(1) \times \U(1)$ invariant one found in \cite{Fischbacher:2010ec} has 4d supergravity modes that could be candidates for marginal deformations. This solution can again be found in the 14-scalar model after setting $a=b=1$ and $\tilde a = \tilde b=0$ in Section~\ref{subsec:14scalar} and the critical point of the potential with value $V = -12 g^2$ is located at
\be
{\bf z} = \Big(\sqrt{2-\sqrt{3}}\, \e^{\frac{i \pi
   }{4}},-\sqrt{2+\sqrt{3}}\, \e^{-\frac{i \pi }{4}},\e^{\frac{i
   \pi }{6}},-\sqrt{2-\sqrt{3}}\, \e^{-\frac{i \pi
   }{4}},-\sqrt{2-\sqrt{3}}\, \e^{-\frac{i \pi }{4}},i,i\Big)\,.
\ee
After computing the mass spectrum of supergravity excitations around this AdS$_4$ vacuum we find the following multiplets in the dual 3d $\mathcal{N}=1$ SCFT
\be
\begin{split}
&A_1[\tfrac52,\tfrac32]\,,\qquad A_1[\tfrac32,\tfrac12]_{\times 2}\,,\\
&L_1[3,1]\,,\qquad 2\times L_1[1+\sqrt{3},1]\,,\qquad 4\times L_1[\tfrac52 ,1]\,,\\
&4\times L_1[2+\sqrt{\tfrac52},\tfrac12]\,,\qquad 4\times L_1[\tfrac52,\tfrac12]\,,\qquad 4\times L_1[\sqrt{\tfrac52} ,\tfrac12]\,,\\
&4\times L_2[1+\sqrt{4\pm\sqrt{15}},0]\,,\quad L_2[2+\sqrt{3},0]\,,\quad L_2[\sqrt{3},0]\,,\quad 4\times L_2[\tfrac32,0]\,,\quad 2\times L_2[1 ,0]\,.\\
\end{split}
\ee
We see that there are no $L_2[2,0]$ multiplets and therefore this critical point does not seem to admit exactly marginal deformations. Curiously we find that there is an $L_1[3,1]$ multiplet which was also present in the examples of 3d $\mathcal{N}=1$ conformal manifolds discussed Sections~\ref{sec:AdS4Sfolds} and \ref{subsec:spec}. It will be interesting to understand the reason for the recurring appearance of this multiplet in the holographic 3d $\mathcal{N}=1$ CFTs discussed here.

\subsection{Conformal manifolds from AdS$_4$ vacua in 10d and 11d supergravity}
\label{subsec:10d11dCMs}

We now discuss two methods by which on can generate family of $\mathcal{N}=1$ supersymmetric AdS$_4$ solutions directly in type IIB and 11d supergravity without appealing to a 4d gauged supergravity truncation.

\subsubsection{Conformal manifolds from Sasaki-Einstein J-folds}
\label{subsubsec:SE5}

In the previous sections we took a four-dimensional point of view to holographically construct and study conformal manifolds preserving 3d $\mathcal N=1$ supersymmetry. In this section we switch gears and study the holographic duals of a class of 3d $\mathcal N=1$ conformal manifolds directly in type IIB supergravity. The supergravity backgrounds generalize the ones described in Section \ref{sec:AdS4Sfolds} and schematically take the form\footnote{When we take the SE$_5$ manifold to be $S^5$ we recover the background discussed in Section \ref{sec:AdS4Sfolds}.}
\begin{eqnarray}
	\text{AdS}_4 \times S^1_\rho \times {\text{SE}}_5\,,
\end{eqnarray}
where there is an $\SL(2,\mathbf Z)_{\text{IIB}}$ monodromy twist on the $S^1_\rho$ and it is fibered over the Sasaki-Einstein manifold (SE$_5$), as we describe explicitly below. The SCFTs dual to these backgrounds arise from S-folding the IR fixed points of four-dimensional $\mathcal N=1$ toric quiver gauge theories.

Our starting point is the type IIB supergravity solutions dual to the $\mathcal N=1$ S-folds described in \cite{Bobev:2019jbi}. The ten-dimensional metric is given by\footnote{We write the metric in Einstein frame, and follow the conventions of \cite{Bobev:2018hbq} for type IIB supergravity. }
\begin{equation}\label{eq:JfoldSE1}
	\rmd s_{10}^2 = \frac{2}{3 g^2} \sqrt{\frac{5}{6}}  \left[  5 \rmd s_{\text{AdS}_4}^2 +  4 \rmd \rho^2 + 6 \rmd s_{\text{SE}_5}^2  \right]\,,
\end{equation}
where $\rho \sim \rho + \Delta \rho$, and $\rmd s_{\text{SE}_5}^2$ is the metric on a Sasaki-Einstein manifold of the form
\begin{equation}\label{eq:JfoldSE2}
	\rmd s_{\text{SE}_5}^2 =  \rmd s_{\text{KE}_4}^2 + \frac{6}{5} (\rmd \psi + \sigma)^2\,,
\end{equation}
with K\"ahler-Einstein base $\rmd s_{\text{KE}_4}^2$. The K\"ahler base is equipped with a K\"ahler $(1,1)$ form, $2J= \rmd \sigma$, and a holomorphic $(2,0)$ form $\Omega$ which obey the relations 
\begin{equation}
	\Omega \wedge \bar \Omega = 2 J\wedge J\,, \quad \text{and} \quad \rmd \Omega = 3\rmi \sigma\wedge \Omega\,.
\end{equation}
We focus now on toric Sasaki-Einstein spaces such that the total isometry group of the manifold is at least $\U(1)_R \times \U(1)_1 \times \U(1)_2$, where $\U(1)_R$ is generated by the Reeb vector present on all Sasakian manifolds. Explicit metrics for two infinite classes of such manifolds, known as $Y^{p,q}$ and $L^{a,b,c}$, were constructed in \cite{Gauntlett:2004yd,Martelli:2005wy,Cvetic:2005ft}, see also \cite{Sparks:2010sn} for a review. 

The five-form flux takes the simple form
\begin{equation}\label{eq:JfoldSE3}
	F_5 = - \frac{24}{g^4} \rmd \psi \w J \w J\,.
\end{equation}
The dilaton-axion is given by
\begin{equation}\label{eq:JfoldSE4}
	\tau = C_0 + \frac{\rmi}{\rme^{\Phi}} = \frac{\cosh(2\rho+\Delta\rho) + i \sinh(\Delta\rho)}{\cosh(2\rho)} \,.
\end{equation}
The three-form fluxes can be compactly written as
\begin{equation}\label{eq:JfoldSE5}
	H_3 + \rmi \rme^{\Phi} F_3  = \frac{2}{g^2}\sqrt{\frac{2}{3 \sinh \Delta \rho}} (\cosh \rho - \rmi \sinh \rho) \left[3 \rme^{\rmi \psi} (\rmd \psi + \sigma) \w \Omega - \rme^{-3\rmi \psi} \rmd \rho \w \bar\Omega \right]\,.
\end{equation}
As is clear from the axio-dilaton and three form fluxes one needs to apply an S-duality transformation along the $S^1_\rho$ direction in order to have consistent  periodicity for all supergravity fields charged with respect to $\SL(2,\mathbf Z)_{\text{IIB}}$. The explicit $\SL(2,\mathbf Z)_{\text{IIB}}$ monodromy transformation used to compactify the $S^1_\rho$ takes the form
\begin{equation}\label{Eq: SL2 element used to compactify}
	\mathcal J_n  = \left(\begin{array}{cc}
		n & -1 \\
		1 & 0
	\end{array}  \right)\,, \qquad \text{with}\quad  n = 2\cosh \Delta\rho\,,
\end{equation}
such that the axio-dilaton and two form gauge potentials transform as\footnote{We follow the conventions of \cite{Polchinski:1998rr}, Section 12.1, for these transformations.}
\begin{equation}
	\tau \rightarrow \frac{1}{-\tau + n}\,, \qquad    \left(\begin{array}{c}
		B_2  \\
		C_2 
	\end{array}  \right) \rightarrow \mathcal J_n \cdot \left(\begin{array}{c}
		B_2  \\
		C_2 
	\end{array}  \right)\,.
\end{equation}
As was discussed in \cite{Bobev:2019jbi} these AdS$_4$ solutions are dual to three-dimensional $\mathcal N=1$ CFTs whose $S^3$ free energy can be computed holographically to equal
\begin{equation}\label{Eq: S3 free energy CFTs}
	\mathcal F_{S^3} = \sqrt{\frac{5^5}{3^6}} \text{arccosh} (n/2) a_{4d}\,,
\end{equation}
where $a_{4d}$ is the central charge of the ``parent'' 4d $\mathcal{N}=1$ quiver gauge theories dual to the type IIB AdS$_5\times {\text{SE}}_5$ Freund-Rubin solutions.

What we will argue now is that these supersymmetric AdS$_4$ backgrounds are not isolated but instead belong to a continuous family whose holographically dual SCFTs share the same value for the $S^3$ free energy. This in turn strongly suggests that the 3d $\mathcal N=1$ CFTs exhibit a conformal manifold. To explicitly construct the continuous family of AdS$_4$ backgrounds we make use of the fact that the toric Sasaki-Einstein manifolds have at least three Killing vectors with an associated $\U(1)_R \times \U(1)_1 \times \U(1)_2$ isometry group. We can then fiber the compact $ S^1_\rho$ direction over these isometry directions and construct new supergravity solutions. One can furthermore show that these fibered backgrounds remain solutions of the equations of motion of type IIB supergravity. To ensure that the AdS$_4$ solutions remain supersymmetric one has to require that the fibration does not affect the Killing spinors of the seed solution in  \eqref{eq:JfoldSE1}-\eqref{eq:JfoldSE5}. Since the Killing spinors on a Sasaki-Einstein manifold are charged with respect to $\U(1)_R$, and uncharged with respect to the $\SL(2,\mathbf Z)$ element in \eqref{Eq: SL2 element used to compactify} it turns out that any mixing of the $\rmd \psi$ direction with $ S^1_\rho$ breaks supersymmetry globally. The Killing spinors are however uncharged with respect to $\U(1)_1 \times \U(1)_2$ so there is no obstruction to mix these two directions with $S^1_\rho$. This is the ten-dimensional manifestation of the supersymmetry preserving deformations described in Section~\ref{flatdirs}.

To show explicitly how the holographic construction of the conformal manifolds works we focus on the simple example where the Sasaki-Einstein manifold is $T^{1,1}$, i.e. the base of the conifold. This is the well-known coset space 
\begin{equation}
	Y^{1,0} = T^{1,1} = \frac{\SU(2)\times\SU(2)}{\U(1)}\,.
\end{equation}
Choosing the coordinates appropriate to the two spheres the metric simply becomes
\begin{equation}
	\rmd s_{Y^{1,0}}^2 = \frac{1}{6} \left[ \rmd \theta_1^2 + \sin^2 \theta_1 \rmd \phi_1^2 +  \rmd \theta_2^2 + \sin^2 \theta_2 \rmd \phi_2^2  \right] + \frac65(\rmd \psi + \sigma)^2\,.
\end{equation}
The K\"ahler one-form, and the holomorphic $(2,0)$ two-form take the explicit form
\begin{equation}
	\sigma = \frac13 \left( \cos\theta_1 \rmd \phi_1 + \cos \theta_2 \rmd \phi_2 \right)\,,\quad \Omega =  \frac16(\rmd \theta_1 - \rmi \sin\theta_1 \rmd \phi_1) \w (\rmd \theta_2 - \rmi \sin\theta_2 \rmd \phi_2)\,.
\end{equation}
The three Killing vectors corresponding to the toric isometries $\U(1)_i$ and $\U(1)_R$ mentioned above are
\begin{equation}
	\partial_{\phi_i}  \,,\quad \text{and} \quad \partial_{\psi}\,,
\end{equation}
To introduce the fibration of the $S^1_\rho$ over the SE$_5$ space we apply the shifts
\begin{equation}\label{eq:phishift}
	\phi_i \rightarrow \hat\phi_i = \phi_i + \chi_i \rho\,.
\end{equation}
Na\"ively one could expect that a similar shift would be possible for the $\psi$ angle, however one can show that such a transformation does not leave the Kililng spinor of this background invariant since it is charged under $\U(1)_R$ and thus breaks supersymmetry. Formally the type IIB supergravity background can be written in the same way as \eqref{eq:JfoldSE1}-\eqref{eq:JfoldSE5} with the replacements
\begin{equation}
	\phi_i \rightarrow \hat \phi_i\,, \quad \sigma(\theta_i,\phi_i) \rightarrow \hat\sigma(\theta_i,\hat\phi_i) \,, \quad J(\theta_i,\phi_i) \rightarrow \hat J(\theta_i,\hat\phi_i) \,, \quad \Omega(\theta_i,\phi_i) \rightarrow \hat\Omega(\theta_i,\hat\phi_i)\,,
\end{equation}
which respect the K\"ahler structure on the base manifold
\begin{equation}
	2\hat J = \rmd \hat \sigma\,, \quad \hat\Omega \w \bar{\hat\Omega} = 2 \hat J\w \hat J\,, \quad \text{and} \quad  \rmd \hat\Omega = 3\rmi \hat\sigma\wedge \hat\Omega\,.
\end{equation}
We have explicitly checked that these new backgrounds are solutions of the equations of motion of type IIB supergravity and, because the spinors are not charged with respect to any of the isometries in the K\"ahler base, they remain supersymmetric as well. Additionally, it is easy to show that the volume form of the background does not change after the deformations in \eqref{eq:phishift}, and thus one concludes that the holographic computation of the dual SCFT $S^3$ free energy remains the same as in equation \eqref{Eq: S3 free energy CFTs}. This is of course compatible with the interpretation that the parameters $\chi_i$ correspond to exactly marginal deformations in the dual SCFT. We end our discussion by emphasizing that in general the coordinate transformation in \eqref{eq:phishift} may not be globally well-defined on the SE$_5$ manifold, especially in the cases of quasi-regular or irregular SE$_5$ manifolds. It will be interesting to understand the global properties of these solutions better.

\subsubsection{TsT transformations}

Another well-known method to construct continuous families of AdS$_4$ solutions of string and M-theory is to use the TsT transformation of Lunin-Maldacena \cite{Lunin:2005jy}. In 10d supergravity the idea is to start with a seed AdS solution that has an internal space with an isometry group that has an $\U(1)^2$ subgroup. Then one can generate a one-parameter family of AdS vacua by applying a series of a T-duality along one of the isometry directions, a coordinate shift along the other one, and then another T-duality on the first direction. If one starts with an 11d supergravity solution one needs an internal space with an isometry group that contains $\U(1)^3$ as a subgroup, then one reduces along one of the isometry directions to type IIA supergravity, applies the TsT procedure there and then uplifts the resulting solution back to 11d. If the seed AdS solution preserves supersymmetry and at least some of the Killing spinors are not charged under the $\U(1)^2$ (or $\U(1)^3$) isometry used in the transformation described above then the whole family of solutions will also preserve supersymmetry. This approach to generate families of AdS solution was used in \cite{Lunin:2005jy} to find the supergravity dual of the so-called marginal $\beta$-deformation of the superpotential of 4d $\mathcal{N}=4$ SYM. In the context of AdS$_4$ solutions the TsT transformation was applied to generate families of AdS$_4$ solutions of 11d and massive IIA supergravity dual to 3d $\mathcal{N}=2$ conformal manifolds, see \cite{Lunin:2005jy,Gauntlett:2005jb,Ahn:2005vc} and \cite{Bobev:2021gza} respectively.

To the best of our knowledge the only examples of holographic duals to 3d $\mathcal{N}=1$ conformal manifolds are the ones constructed in \cite{Gauntlett:2005jb} based on Freund-Rubin solutions with weak G$_2$ or tri-Sasakian internal manifolds. Here we would like to point out that one can also apply the TsT procedure to supergravity solutions that are not of the Freund-Rubin type and can be explicitly constructed by uplifting AdS$_4$ vacua of gauged supergravity theories. We will only discuss examples based on the known supersymmetric vacua of the gauged supergravity theories summarized in Table~\ref{tableofgaugings}.

We start by noting that there are three known AdS$_4$ $\mathcal{N}=1$ supersymmetric vacua of the $\SO(8)$ maximal gauged supergravity, see \cite{Comsa:2019rcz,Bobev:2019dik}. The isometry group of the internal manifolds corresponding to these solutions are G$_2$, $\SO(3)$ and $\U(1)\times \U(1)$, respectively. Since these are 11d supergravity vacua we conclude that we do not have a large enough isometry group to apply the TsT method to these solutions. For the $[\SO(6)\times \SO(2)]\ltimes \R^{12}$ gauged supergravity the only known supersymmetric AdS$_4$ solutions are the ones presented in Section~\ref{sec:AdS4FFG}. Both of them uplift to type IIB supergravity solutions that do not have an isometry group that contains $\U(1)^2$ so one cannot use the TsT transformation to generate additional continuous deformations of these backgrounds.

The situation is more interesting for the ISO(7) gauged supergravity. There are 7 known supersymmetric AdS$_4$ solutions in this theory, see \cite{Bobev:2020qev} for a recent summary. Four of these solutions admit uplifts to massive IIA supergravity that can be deformed by the TsT transformation. For the $\mathcal{N}=2$ AdS$_4$ solutions with $\SU(3) \times \U(1)$ symmetry the TsT transformations were recently presented in \cite{Bobev:2021gza}. The G$_2$ invariant $\mathcal{N}=1$  AdS$_4$ solution found in \cite{Behrndt:2004km,Borghese:2012qm} as well as the $\SU(3)$ invariant $\mathcal{N}=1$ solution of \cite{Guarino:2015qaa} should both allow for 1-parameter family of continuous deformations in IIA supergravity constructed by applying the TsT procedure.\footnote{Note also that that the 10d uplift of the $\SU(3)$ $\mathcal{N}=1$ solution has the explicit K\"ahler-Einstein metric on $\mathbb{CP}^2$ and is based on the Sasaki-Einstein structure of $S^5$  \cite{Varela:2015uca}. This leads to a generalization of this solution based on arbitrary Sasaki-Einstein spaces. If these solutions have at least $\U(1)^2$ isometry then they can also be subjected to the TsT transformation.} The final example is the $\SO(3)_F\times \SO(3)_R$ invariant $\mathcal{N}=3$ vacuum of the ISO(7) gauged 4d supergravity found in \cite{Gallerati:2014xra} is presented in \cite{Pang:2015vna}. At first sight it may seem that applying the TsT transformation to this solution will break supersymmetry completely. However since there are 3 Killing spinors in the $\mathbf{3}$ of the $\SO(3)_R$ R-symmetry there will be a singlet spinor after the TsT transformation which in turn will result in a family of $\mathcal{N}=1$ AdS$_4$ vacua with $\U(1) \times \U(1)$ flavor symmetry.

Finally we note that in the  $[\SO(6)\times \SO(1,1)]\ltimes \mathbf{R}^{12}$ gauged supergravity there is only a single supersymmetric AdS$_4$ solution which can be uplifted to IIB supergravity and subjected to the TsT transformation. This is the $\mathcal{N}=1$ solution discussed in Section~\ref{sec:AdS4Sfolds}. Its uplift leads to an isometry group that is either $\SU(3)$, for $\chi_i = 0$, or a smaller subgroup for non-zero $\chi_i$. For any value of $\chi_i$ the uplifted solution can be deformed by the TsT transformation while preserving supersymmetry. The same applies also to the generalization of this solution based on Sasaki-Einstein manifolds discussed in Section~\ref{subsubsec:SE5}. This suggests that these supergravity solutions are dual to 3d $\mathcal{N}=1$ conformal manifolds spanning three real dimensions - two marginal deformations corresponding to the 4d supergravity parameters $\chi_i$ and one arising from the TsT transformations.

It will of course be interesting to construct explicitly the TsT deformations of the supergravity solutions discussed above and compute (part of) the KK spectrum as a function of the deformation parameters. This could shed some light on the dual CFT descriptions of these models.

\section{Discussion}
\label{sec:discussion}

In this paper we constructed and explored some explicit examples of families of supersymmetric AdS$_4$ supergravity solutions which are holographically dual to 3d $\mathcal{N}=1$ SCFTs with conformal manifolds. The main examples we focused on arise in type IIB string theory from the worldvolume dynamics of $N$ D3-branes wrapped on $S^1$ with an S-duality twist. In these cases we showed that the 3d $\mathcal{N}=1$ conformal manifolds are compact and computed the spectrum of low-lying conformal dimensions as functions of the exactly marginal coupling. We also calculated the free energy on $S^3$ in the planar limit of these SCFTs. Our results lead to some natural questions and avenues for future work which we discus below.

It should be possible to understand additional properties of the new 4d gauged supergravity solutions we constructed by uplifting them to solutions of type IIB supergravity or by calculating the full spectrum of KK modes.\footnote{The $\SO(3)$ invariant solution was uplifted to type IIB supergravity in \cite{Berman:2021ynm}.} The latter can be evaluated by using techniques from exceptional field theory as was recently done for holographic 3d $\mathcal{N}=2$ conformal manifolds in \cite{Giambrone:2021zvp,Cesaro:2021tna}. It will also be interesting to find other explicit examples, either in 4d or 10d/11d supergravity, of families of supersymmetric AdS$_4$ solutions dual to 3d $\mathcal{N}=1$ SCFTs with conformal manifolds. In Section~\ref{subsec:10d11dCMs} we outlined how some examples of these solutions can be constructed but it is desirable to develop other techniques for finding such backgrounds. 

An alternative point of view on AdS$_4$ solutions arising from type IIB string theory with an S-fold is offered by 5d maximal $\SO(6)$ gauged supergravity, see \cite{Bobev:2019jbi,Bobev:2020fon,Arav:2020asu,Arav:2020obl,Arav:2021tpk} for some recent discussions on such constructions. It is then natural to wonder how the axionic scalars $\chi$ that lead to the marginal deformations are realized in 5d. We have explicitly verified that these 4d scalar fields arise from a Wilson loop in a representation of the Cartan subgroup of the $\SO(6)$ gauge group along the $S^1_{\rho}$ direction involved in the S-fold procedure. The masslessness of the 4d scalar field $\chi$ can then be understood as originating from the fact that the 5d $\SO(6)$ gauge field is massless.\footnote{This point was also recently emphasized in \cite{Berman:2021ynm}.}

The results summarized above clearly suggest that there are new classes of 3d $\mathcal{N}=1$ SCFTs with exactly marginal couplings that arise from D3-branes in string theory. It will of course be very interesting to find purely field theoretic mechanism to understand the dynamics of these theories. The supergravity construction indicates that the 3d SCFTs should be related to the low-energy limit of $\mathcal{N}=4$ SYM on $\mathbf{R}^{1,2}\times S^1_{\rho}$ with a twist by a hyperbolic or an elliptic element of the SL$(2,\mathbf{Z})$ S-duality group. A natural guess is that these 3d SCFTs are in some way related to the strongly coupled  $T[\U(N)]$ 3d $\mathcal{N}=4$ SCFT of Gaiotto-Witten \cite{Gaiotto:2008ak} with some additional deformations that break the supersymmetry to $\mathcal{N}=1$. The exactly marginal operators in these  3d $\mathcal{N}=1$ SCFTs should then arise from Wilson lines of the unbroken global symmetry of the Abelian subgroup of the $\SO(6)$ R-symmetry preserved by the construction. It will be most interesting to develop technical tools to study these types of models quantitatively and compare with the holographic description we presented above.

Another important open question, both in field and in string theory, is to understand whether the continuous families of solutions we have found are artefacts of the large $N$ approximation.\footnote{As for example was found in a simpler setting in \cite{Aharony:2011jz,Aharony:2019mbc}, where the beta function of a $\phi^6$ coupling was studied in a large $N$ limit of a vector model coupled to a gauge field with a Chern-Simons term.} On the 4d supergravity side it is known that both the superpotential and the K\"ahler potential can in principle receive corrections. With $\mathcal{N}=1$ supersymmetry the superpotential should not receive perturbative corrections, however the K\"ahler potential can receive both perturbative and non-perturbative corrections, see for example \cite{Aharony:2012jf} for a summary in the AdS$_4$ context. It is of course very interesting to understand whether the continuous family of AdS$_4$ vacua we discussed here is affected by such corrections from string theory effects. In \cite{deAlwis:2013jaa} general properties of the moduli spaces of  AdS$_4$ $\mathcal{N}=1$ supergravity solutions were studied, however no explicit top-down examples were discussed. We do not have in mind any particular string theory mechanism that will lift the flat directions we have discovered and in fact are not aware of any examples in string theory where continuous families of supersymmetric AdS$_4$ vacua of supergravity are destroyed by quantum corrections.\footnote{Recently some additional arguments for the non-perturbative stability of continuous families of AdS$_4$ vacua very similar to the ones discussed here were summarized in \cite{Giambrone:2021wsm}.} Note also that the candidate exactly marginal deformations discussed above should be dual to single-trace operators in the dual SCFT and thus should not suffer from the double-trace instability discussed in \cite{Berkooz:1998qp}. This discussion leads to the very interesting prospect that the 3d $\mathcal{N}=1$ conformal manifolds we have identified in this work are present also at finite $N$. We are then presented with the important task of identifying the CFT mechanism for the existence of exactly marginal operators in strongly coupled 3d CFTs with only $\mathcal{N}=1$ supersymmetry.\footnote{Since the 3d $\mathcal{N}=1$ SCFTs discussed in Section~\ref{sec:AdS4Sfolds} and \ref{sec:AdS4FFG} should arise from placing 4d $\mathcal{N}=4$ SYM on $S^1$ and performing an S-duality twist, perhaps the mechanism for appearance of exactly marginal couplings in 3d BCFTs studied in  \cite{DiPietro:2019hqe} may play a role.}

We hope that our work will stimulate the study of 3d $\mathcal{N}=1$ conformal manifolds and that further examples, both in field theory and holography, will be found and analyzed in more detail.

\bigskip

\noindent\textbf{Acknowledgements }

\medskip

\noindent We are grateful to Ofer Aharony, Francesco Benini, Krzysztof Pilch, and Thibeau Wouters for a number of useful discussions. We would also like to thank Thomas Fischbacher for illuminating discussions and early collaboration on some of the results presented here. The work of NB is supported in part by an Odysseus grant G0F9516N from the FWO. FFG is supported by the University of Iceland Recruitment Fund. The work of JvM was partially supported by a doctoral fellowship from the Research Foundation - Flanders (FWO). NB and JvM are also partially supported by the KU Leuven C1 grant ZKD1118 C16/16/005. JvM is supported by the MIUR-PRIN contract 2015 MP2CX4, and by the ERC-COG grant NP-QFT no.~864583, ``Non-perturbative dynamics of quantum fields: from new deconfined phases of matter to quantum black holes''. 

\begin{appendices}

\appendix

\section{Non-supersymmetric AdS$_4$ solutions}
\label{app:nonsusyAdS4}

In this appendix we present the non-supersymmetric AdS$_4$ solutions we have found of the 14 scalar model with $[\SO(6)\times \SO(2)]\ltimes \R^{12}$ gauging. We found 15 solutions in total, two of them are the ${\cal N}=1$ supersymmetric solutions discussed in Section~\ref{sec:AdS4FFG}. In addition, we have 13 non-supersymmetric solutions. The position of the solutions is given as approximate values for the 7 $z_i$ scalars. We have determined the continuous symmetry for all solutions, and found the number of flat directions of the potential for each solution. For all solutions with a continuous global symmetry we find that the procedure described in Section~\ref{flatdirs} leads to a continuous family of solutions along which the symmetry is broken to its Cartan subalgebra. For all solutions we numerically computed the scalar masses for the 70 scalars of the 4d supergravity theory as a function of the marginal parameters in order to check for the BF stability of the AdS$_4$ vacua. We find that all non-supersymmetric solutions are BF unstable. This information is collected in Table \ref{nonsusytable}. Below we list the position of the 13 non-supersymmetric solutions.  
\begin{table}[h!]
\begin{center}
\begin{tabular}{@{\extracolsep{10 pt}}l c c c c}
\toprule
Potential& SUSY & Cont. symmetry &  stability & \# flat directions \\
\midrule
\hyperref[752908]{$-7.52908\, g^2$} & ${\cal N}=1$ & $\SO(3)$		& S & 1\\
\hyperref[758297]{$-7.58297\, g^2$} & ${\cal N}=1$ & $\U(1)$			& S & 1\\
\midrule
\hyperref[519615]{$-5.19615\, g^2$} & ${\cal N}=0$ & $\SO(4)$ 		& U & 2\\
\hyperref[524301]{$-5.24301\, g^2$} & ${\cal N}=0$ & $\U(1)$			& U & 1\\
\hyperref[554078]{$-5.54078\, g^2$} & ${\cal N}=0$ & $\emptyset$	& U & 0\\ 
\hyperref[554875]{$-5.54875\, g^2$} & ${\cal N}=0$ & $\emptyset$	& U & 0\\
\hyperref[558415]{$-5.58415\, g^2$} & ${\cal N}=0$ & $\U(1)$			& U & 1\\
\hyperref[560429]{$-5.60429\, g^2$} & ${\cal N}=0$ & $\U(1)$			& U & 1\\
\hyperref[565862]{$-5.65862\, g^2$} & ${\cal N}=0$ & $\U(1)$			& U & 1\\
\hyperref[573131]{$-5.73131\, g^2$} & ${\cal N}=0$ & $\U(1)$			& U & 1\\
\hyperref[579114]{$-5.79114\, g^2$} & ${\cal N}=0$ & $\SO(3)$		& U & 1\\
\hyperref[582843]{$-5.82843\, g^2$} & ${\cal N}=0$ & $\emptyset$	& U & 0\\
\hyperref[622410]{$-6.22410\, g^2$} & ${\cal N}=0$ & $\emptyset$	& U & 0\\
\hyperref[793185]{$-7.93185\, g^2$} & ${\cal N}=0$ & $\emptyset$	& U & 0\\
\hyperref[987723]{$-9.87723\, g^2$} & ${\cal N}=0$ & $\emptyset$	& U & 0\\
\bottomrule
\end{tabular}
\caption{The 15 AdS$_4$ solutions of the 14-scalar model with $[\SO(6)\times \SO(2)]\ltimes \R^{12}$ gauging. The specified continuous symmetry refers to the maximally symmetric point along a possible family of critical points sharing the same value of the potential.}
\label{nonsusytable}
\end{center}
\end{table}

\paragraph{$V=-5.19615\, g^2 = -3\sqrt{3}\,g^2$}
\label{519615}
\be
\begin{split}
z_1 &=z_2=z_3= i/\sqrt{3} \,,\\
z_4 &=z_5=z_7=-\bar{z}_6=(1+i)/\sqrt{2} \,.
\end{split}
\ee

\paragraph{$V=-5.24301\, g^2$}
\label{524301}
\be
\begin{split}
z_1 &\approx 0.0752+0.86328 i\,,\\
z_2 &=z_3 \approx 0.55825 i\,,\\
z_4 &=-\bar{z}_7 \approx0.77717+0.62929 i\,,\\
z_5 &\approx-1.0622+0.92739 i\,,\\
z_6 &\approx-0.53422+0.46642 i\,.
\end{split}
\ee

\paragraph{$V=-5.54078\, g^2$}
\label{554078}
\be
\begin{split}
z_1 &\approx 0.22956+0.81359 i\,,\\
z_2 &\approx -0.10769+0.43692 i\,,\\
z_3 &\approx 0.22408+0.66626 i\,,\\
z_4 &\approx -0.22712+0.88864 i\,,\\
z_5 &\approx 0.34004+1.6207 i\,,\\
z_6 &\approx-0.21853+0.72191 i\,,\\
z_7 &\approx 0.86772+0.227 i\,.
\end{split}
\ee

\paragraph{$V=-5.54875\, g^2$}
\label{554875}
\be
\begin{split}
z_1 &= z_3 \approx -0.17712+0.5362 i\,,\\
z_2 &\approx -0.24442+0.82564 i\,,\\
z_4 &\approx 0.18395+0.62614 i\,,\\
z_5 &\approx 0.96844+0.24923 i\,,\\
z_6 &\approx 0.43192+1.4702 i\,,\\
z_7 &\approx 0.25748+0.96628 i\,.
\end{split}
\ee

\paragraph{$V=-5.58415\, g^2$}
\label{558415}
\be
\begin{split}
z_1 &-\bar{z}_3\approx -0.26788+0.8214 i\,,\\
z_2 &\approx 0.1467+0.34994 i\,,\\
z_4 &\approx -0.84264+0.53847 i\,,\\
z_5 &-\bar{z}_7\approx 0.82159+0.57007 i\,,\\
z_6 &\approx -0.56005+0.82846 i\,.
\end{split}
\ee

\paragraph{$V=-5.60429\, g^2$}
\label{560429}
\be
\begin{split}
z_1 &\approx -0.32226+0.71107 i\,,\\
z_2 &\approx 0.40798 i\,,\\
z_3 &\approx 0.87161 i\,,\\
z_4 &\approx 1.1192 i\,,\\
z_5 &\approx -1.1615+0.29256 i\,,\\
z_6 &\approx -0.50926+1.2306 i\,,\\
z_7 &\approx 0.52388 i\,.
\end{split}
\ee

\paragraph{$V=-5.65862\, g^2$}
\label{565862}
\be
\begin{split}
z_1 &\approx +0.35319 i\,,\\
z_2 &\approx +0.79582 i\,,\\
z_3 &\approx -0.40714+0.94357 i\,,\\
z_4 &\approx -0.81032+0.1898 i\,,\\
z_5 &\approx -0.35266+0.87056 i\,,\\
z_6 &\approx +0.77015 i\,,\\
z_7 &\approx +1.7353 i\,.
\end{split}
\ee

\paragraph{$V=-5.73131\, g^2$}
\label{573131}
\be
\begin{split}
z_1 &=-\bar{z}_3\approx 0.26371+0.67265 i\,,\\
z_2 &\approx 0.23082+0.48246 i\,,\\
z_4 &\approx -0.42581+0.78857 i\,,\\
z_5 &=-\bar{z}_7\approx 0.97052+0.61309 i\,,\\
z_6 &\approx -0.67594+0.43339 i\,.
\end{split}
\ee

\paragraph{$V=-5.79114\, g^2$}
\label{579114}
\be
\begin{split}
z_1 &=z_2=-\bar{z}_3 \approx 0.27172+0.57642 i\,,\\
z_4 &=z_5=-\bar{z}_7\approx \,,0.37759+0.92597 i\\
z_6 &\approx -0.96834+0.24963 i\,.
\end{split}
\ee

\paragraph{$V=-5.82843\, g^2$}
\label{582843}
\be
\begin{split}
z_1 &\approx +1.0192 i\,,\\
z_2 &\approx +0.28738 i\,,\\
z_3 &\approx 0.5412+ i\,,\\
z_4 &\approx 0.45692+0.88951 i\,,\\
z_5 &\approx -0.98122+0.19289 i\,,\\
z_6 &=z_7= i\,.
\end{split}
\ee

\paragraph{$V=-6.22410\, g^2$}
\label{622410}
\be
\begin{split}
z_1 &\approx -0.392+1.2574 i\,,\\
z_2 &=-\bar{z}_3\approx -0.25325+0.54012 i\,,\\
z_4 &\approx 0.81491+2.1165 i\,,\\
z_5 &\approx 0.28599+0.95823 i\,,\\
z_6 &\approx 0.98756+0.15726 i\,,\\
z_7 &\approx -0.15842+0.41147 i\,.
\end{split}
\ee

\paragraph{$V=-7.93185\, g^2$}
\label{793185}
\be
\begin{split}
z_1 &\approx 0.47344+0.31763 i\,,\\
z_2 &\approx 0.25978+0.4808 i\,,\\
z_3 &\approx 0.54128+0.63009 i\,,\\
z_4 &\approx -0.73901+0.87915 i\,,\\
z_5 &\approx -0.26803+1.1537 i\,,\\
z_6 &\approx -0.85698+1.4975 i\,,\\
z_7 &\approx 0.58745+0.19852 i\,.
\end{split}
\ee

\paragraph{$V=-9.87723\, g^2$}
\label{987723}
\be
\begin{split}
z_1 &\approx -0.88319+0.53398 i\,,\\
z_2 &\approx -0.48195+0.63469 i\,,\\
z_3 &\approx 0.28568+0.30692 i\,,\\
z_4 &\approx -0.46442+0.34271 i\,,\\
z_5 &\approx -0.29589+0.56915 i\,,\\
z_6 &\approx 0.61876+1.1477 i\,,\\
z_7 &\approx -1.4222+0.26466 i\,.
\end{split}
\ee


\end{appendices}

\bibliography{3dN=1CMs-AdS4}
\bibliographystyle{utphys}

\end{document}